\documentclass[10pt, conference, compsocconf]{IEEEtran}

\usepackage{cite}

\usepackage[main=english,french]{babel}
\usepackage[utf8]{inputenc}
\usepackage{comment}
\usepackage{listings}
\usepackage{url}
\usepackage{makecell}
\usepackage{multirow}
\usepackage{amsmath}
\usepackage{amssymb}

\usepackage{hyperref}
\usepackage[pdftex]{graphicx}
\DeclareGraphicsExtensions{.png}

\usepackage{xspace}
\addto\captionsenglish{%
}

\newcommand{\domain}{\mathcal{M}}
\newcommand{\reeb}{\mathcal{R}(f)}
\newcommand{\mt}{\mathcal{T}(f)}
\newcommand{\ct}{\mathcal{C}(f)}
\newcommand{\joinTree}{\mathcal{T}^-(f)}
\newcommand{\splitTree}{\mathcal{T}^+(f)}
\newcommand{\range}{\mathbb{R}}

\newcommand{\sub}[1]{f^{-1}_{-\infty}(#1)}
\newcommand{\sur}[1]{f^{-1}_{+\infty}(#1)}
\newcommand{\st}[1]{St(#1)}

\newcommand{\lk}[1]{Lk(#1)}
\newcommand{\lkminus}[1]{Lk^-(#1)}
\newcommand{\lkminusV}[1]{Lk^-_0(#1)}
\newcommand{\lkplus}[1]{Lk^+(#1)}
\newcommand{\lkplusV}[1]{Lk^+_0(#1)}
\newcommand{\vertexNumber}{|\sigma_0|}
\newcommand{\edgeNumber}{|\sigma_1|}
\newcommand{\priorityQueue}{\mathcal{Q}}
\newcommand{\combineQueue}{\mathcal{Q}}

\newcommand{\vcaption}[1]{%
\caption{#1}
}

\usepackage{algorithm}
\usepackage{algpseudocode}

\floatname{algorithm}{Alg.}

\usepackage{tabu}
\usepackage{booktabs}                  
\usepackage{multirow}
\usepackage[table,xcdraw]{xcolor}
\usepackage[singlelinecheck=false,font=footnotesize]{caption}

\let\labelindent\relax
\usepackage{enumitem}
\setlist{leftmargin=7mm}

\newcommand{\Charles}[1]{#1}
\newcommand{\CharlesV}[1]{#1}
\newcommand{\Pierre}[1]{#1}

\newcommand{\Disable}[1]{\textcolor{gray}{#1}}

\usepackage[caption=false,font=footnotesize,labelfont=sf,textfont=sf]{subfig}

\begin{document}

\title{Task-based Augmented Contour Trees with Fibonacci Heaps}


\author{\IEEEauthorblockN{
    C. Gueunet\IEEEauthorrefmark{1}\IEEEauthorrefmark{3},
    P. Fortin\IEEEauthorrefmark{1}\IEEEauthorrefmark{2},
    J. Jomier\IEEEauthorrefmark{3} and
    J. Tierny\IEEEauthorrefmark{1}
  }
  \IEEEauthorblockA{\IEEEauthorrefmark{1}
    Sorbonne Université,  CNRS, Laboratoire d'Informatique de Paris 6, LIP6, F-75005 Paris,
    France}
  \IEEEauthorblockA{\IEEEauthorrefmark{2}
    Université de Lille, CNRS, Centrale Lille, CRIStAL UMR 9189, F-59000 Lille,
    France}
  \IEEEauthorblockA{\IEEEauthorrefmark{3}
    Kitware SAS, Villeurbanne,
    France}
    Email: \{Charles.Gueunet,\,Julien.Jomier\}@kitware.com \{Pierre.Fortin,\,Julien.Tierny\}@sorbonne-universite.fr
}


\maketitle

\begin{abstract}
   This paper presents a new algorithm for the fast, shared memory, multi-core
   computation of augmented contour trees on triangulations. In contrast to
   most existing parallel algorithms our technique computes \emph{augmented}
   trees, enabling the full extent of contour tree based applications including
   data segmentation. Our approach completely revisits the traditional,
   sequential contour tree algorithm to re-formulate all the steps of the
   computation as a set of independent local tasks. This includes a new
   computation procedure based on Fibonacci heaps for the join and split trees,
   two intermediate data structures used to compute the contour tree, whose
   constructions are efficiently carried out concurrently thanks to the dynamic
   scheduling of task parallelism. We also introduce a new parallel algorithm
   for the combination of these two trees into the output global contour tree.
   Overall, this results in superior time performance in practice, both in
   sequential and in parallel thanks to the OpenMP task runtime. We report
   performance numbers that compare our approach to reference sequential and
   multi-threaded implementations for the computation of augmented merge and
   contour trees. These experiments demonstrate the run-time efficiency of our
   approach and its scalability on common workstations. We demonstrate the
   utility of our approach in data segmentation applications.
\end{abstract}



%
\IEEEpeerreviewmaketitle%

\section{Introduction}\label{section:introduction}



\IEEEPARstart{A}{s} scientific data sets become more intricate and larger in
size, advanced data analysis algorithms are needed for their efficient
visualization and interactive exploration. For scalar field visualization,
topological data analysis techniques~\cite{edelsbrunner09,
pascucciTOPOVIS10, heine16, tiernyBook18} have shown to be practical
solutions in various contexts by enabling the concise and complete capture of
the structure of the input data into high-level topological abstractions such
as merge trees~\cite{bremerTVCG11, morozov13, smirnov17}, contour
trees~\cite{boyell63, de1997trekking, tarasov98, carr00}, Reeb
graphs~\cite{reeb46, shinagawa91, pascucci07, biasotti08, tiernyVis09}, or
Morse-Smale complexes~\cite{gyulassy08, robins11, WeissIFF13, guylassyVIS14,
Defl15}. Such topological abstractions are fundamental data-structures that
enable the development of advanced data analysis, exploration and visualization
techniques, including for instance: small seed set extraction for fast
isosurface traversal~\cite{vanKreveld97, carr04}, feature
tracking~\cite{sohn06}, data simplification~\cite{tiernyVIS12},
summarization~\cite{WeberBP07, pascuccimulti} and
compression~\cite{solerPV18}, transfer function design~\cite{weber07},
similarity estimation~\cite{hilaga:sig:2001, tiernyCGF09, thomas14}, or
geometry processing~\cite{dong:tog:2006, tiernyTVCG11, vintescu17}. Moreover,
their ability to capture the features of interest in scalar data  in a generic,
robust and multi-scale manner has contributed to their popularity in a variety
of applications, including turbulent combustion~\cite{laneyVIS06,
bremerTVCG11, gyulassyEV14}, computational fluid
dynamics~\cite{kastenTVCG11, fangChen13}, material
sciences~\cite{gyulassy07, gyulassyVIS15, visCon2016submit, Lukasczyk17},
chemistry~\cite{chemistryVIS14}, and astrophysics~\cite{sousbie11,
shivashankar2016felix, rosen17}.

However, as computational resources and acquisition devices improve, the
resolution of the geometrical domains on which scalar fields are defined also
increases. This increase \Charles{in the input size} yields several technical
challenges for topological data analysis, including that of computation time
efficiency, which is a critical criterion in the context of interactive data
analysis and exploration, where the responsiveness of the system to user
queries is of paramount importance. Thus, to enable truly interactive
exploration sessions, highly  efficient algorithms are required for the
computation of topological abstractions. A natural direction towards the
improvement of the time efficiency of topological data analysis  is
parallelism, as all commodity hardware (from tablet devices to high-end
workstations) now embeds processors with multiple cores. However, most
topological analysis algorithms are originally intrinsically sequential as they
often require a global view of the data. Thus, in this work, we focus on
parallel approaches for topological methods with the specific target of
improving run times in an interactive environment, where the response time of a
system should be as low as possible.

\Charles{In this paper we focus on the contour tree, which is} a fundamental
topology-based data structure in scalar field visualization. Several algorithms
have been proposed for its parallel computation~\cite{maadasamy12, acharya15,
carr16}. However, these algorithms only compute \emph{non-augmented} contour
trees~\cite{carr00}, which only represent the connectivity evolution of the
sub-level sets, and not the corresponding data-segmentation (i.e.\ the arcs are
not augmented with regular vertices). While such non-augmented trees enable
some of the traditional visualization applications of the contour tree, they do
not enable them all. For instance, they do not readily support topology based
data segmentation. Moreover, fully augmenting in a post-process non-augmented
trees is a non trivial task, for which no linear-time algorithm has ever been
documented to our knowledge.

\Charles{The} new algorithm \Charles{we present here allows} for the efficient
computation of augmented contour trees of scalar data on triangulations. Such a
tree augmentation makes our output data-structures generic application-wise and
enables the full extent of contour tree based applications, including data
segmentation. Our approach completely revisits the traditional, sequential
contour tree algorithm to re-formulate all the steps of the computation as a
set of local tasks that are as independent as possible. This includes a new
computation procedure based on Fibonacci heaps for the join and split trees,
two intermediate data structures used to compute the contour tree, whose
constructions are efficiently carried out concurrently thanks to the dynamic
scheduling of task parallelism.

We also introduce a new parallel algorithm for the combination of these two
trees into the output global contour tree. This results in a computation with
superior time performance in practice, in sequential as well as in parallel,
thanks to the OpenMP task runtime, on multi-core CPUs with shared memory
(typically found on the workstations used for interactive data analysis and
visualization)

Extensive experiments on a variety of real-life data sets demonstrate the
practical superiority of our approach in terms of time performance in
comparison to sequential~\cite{libtourte} and
parallel~\cite{gueunet2016contour} reference implementations, both for
augmented merge and contour tree computations. We illustrate the utility of our
approach with specific use cases for the interactive exploration of hierarchies
of topology-based data segmentations that were enabled by our algorithm. We
also provide a lightweight VTK-based C++ reference implementation of our
approach for reproducibility purposes.


\subsection{Related work}%
\label{section:RelatedWork}

The contour tree~\cite{boyell63}, a tree that contracts connected components of
\emph{level sets} to points (formally defined in \autoref{section:Background}),
is closely related to the notion of merge tree, which contracts connected
components of \emph{sub-level sets} to points on simply connected domains. As
shown by Tarasov and Vyali~\cite{tarasov98} and later generalized by Carr et
al.~\cite{carr00} in arbitrary dimension, the contour tree can be efficiently
computed by combining with a simple linear-time traversal the merge trees of
the input function and of its opposite (called the join and split trees, see
\autoref{section:Background}). Due to this tight relation, merge and contour
trees have often been investigated jointly in the literature.

A simple sequential algorithm, based on a union-find
data-structure~\cite{cormen09}, is typically used for merge tree
computation~\cite{tarasov98, carr00}. It is both simple to implement,
relatively efficient in practice and with optimal time complexity. In
particular, this algorithm allows for the computation of both augmented and
non-augmented merge trees. An open source reference implementation
(\emph{libtourtre}~\cite{libtourte}) of this algorithm is provided by Scott
Dillard. Chiang et al.~\cite{chiang05}  presented an output-sensitive approach,
based on a new algorithm for the computation of non-augmented merge trees using
monotone paths, where the arcs of the merge trees were evaluated by considering
monotone paths connecting the critical points of the input scalar field. Among
the popular applications of the contour tree, interactive data segmentation is
particularly prominent with usages in a variety of domains, as mentioned in the
introduction.
However, these applications of the contour tree to data segmentation require
the \emph{augmented} contour tree as they rely on the identification of the
sets of regular vertices mapping to each arc of the contour tree to extract
regions of interest.

Among the approaches which addressed the time performance improvement of
contour tree computation through shared-memory parallelism, only a few of them
rely directly on the original merge tree computation algorithm~\cite{tarasov98,
carr00}. This algorithm is then used within partitions of the mesh resulting
from a static decomposition on the CPU cores, by either dividing the
geometrical domain~\cite{pascucci03} or the  scalar range~\cite{gueunet2016contour}.
This leads in both cases to extra  computation (with respect to the sequential
mono-partition computation) at the partition boundaries when joining results
from different partitions. This can also lead to load imbalance among the
different partitions~\cite{gueunet2016contour}.

In contrast, most approaches addressing shared-memory parallel contour tree
computation actually focused on revisiting the merge tree sub-procedure, as it
constitutes the main computational bottleneck overall (see
\autoref{section:PerformancesContour}). Maadasamy et al.~\cite{maadasamy12}
introduced a multi-threaded variant of the output-sensitive algorithm by Chiang
et al.~\cite{chiang05}, which results in good scaling performances on
tetrahedral meshes. However, we note that, in practice, the sequential version
of this algorithm is up to three times slower than the reference implementation
(\emph{libtourtre}~\cite{libtourte}, see Tab. 1 in~\cite{maadasamy12}). This
only yields eventually speedups between 1.6 and 2.8 with regard to
libtourtre~\cite{libtourte} on a 8-core CPU~\cite{maadasamy12} (20\% and 35\%
parallel efficiency respectively). We suspect that these moderate speedups over
libtourtre are due to the lack of efficiency of the sequential algorithm based
on monotone paths by Chiang et al.~\cite{chiang05} in comparison to that of
Carr et al.~\cite{carr00}. Indeed, from our experience, although the extraction
of the critical points of the field is a local operation~\cite{banchoff70}, we
found in practice that its overall computation time is often larger than that
of the contour tree itself. Moreover, this algorithm triggers monotone path
computations for each saddle point~\cite{chiang05}, even if it does not yield
branching in the join or split trees (which induces unnecessary computations).
Finally, since it connects critical points through monotone paths, this
algorithm does not visit all the vertices of the input mesh. Thus it cannot
produce an augmented merge tree and consequently cannot support merge tree
based data segmentation.
\Charles{Carr et al.~\cite{carr16} presented a new algorithm available in
the VTK-m library~\cite{vtkm}. This approach is based on massive,
fine-grained (one thread  per input vertex), data parallelism and is specially designed for many-core
architectures
(like GPUs). However, existing implementations only support non-augmented trees
and experiments have only been documented in 2D \cite{carr16}.}
\Charles{In contrast, our approach 
  is based on
coarse-grained parallelism (one thread at a time per output arc) for
multi-core architectures and benefits
from the flexibility of the dynamic load balancing induced by
the task runtime.}
Smirnov et al.~\cite{smirnov17} described a new
data-structure for computing the same information as the merge tree. This
structure can be computed in parallel by using an algorithm close to Kruskal's
algorithm. However, documented experiments report that this algorithm needs at
least 4 threads to be more efficient than a version optimized for a sequential
usage (without atomic variables). Moreover, it has a maximum parallel
efficiency of 18.4\% compared to this optimized sequential version on 32 CPU
cores. Acharya and Natarajan~\cite{acharya15} specialized and improved
monotone-path based computations for the special case of regular grids. Rosen
et al.\ also presented a hybrid CPU-GPU approach for regular
grids~\cite{rosen2017hybrid}. In this work, we focus on triangulations because
of the genericity of this representation: any mesh can be decomposed into a
valid triangulation and regular grids can be implicitly triangulated with no
memory overhead~\cite{ttk}.

To compute the contour tree, two intermediate data-structures, the join and
split trees, need to be \emph{combined} into the global output contour tree
\cite{carr00}. Regarding this combination step, the existing parallel methods
to contour tree computation use almost directly the reference sequential
algorithm~\cite{carr00}. Some parallel attempts for this combination step have
been described in~\cite{carr16,maadasamy12, acharya15}, but no experimental
result \Pierre{concerning this step} has been documented.

Morozov and Weber~\cite{morozov13, morozov13b} and Landge et
al.~\cite{landge14} presented three approaches for merge and contour tree-based
visualization in a distributed environment, with minimal inter-node
communications. However, these approaches focus more on the reduction of the
communication between the processes than on the efficient computation on a
single shared memory node as we do here with the target of an efficient
interactive exploration in mind.

\subsection{Contributions}%
\label{ssection:Contributions}

This paper is an extended version of a conference paper~\cite{gueunet2017ftm},
which made the following contributions.
\begin{enumerate}
   \item{\textbf{A new local algorithm based on Fibonacci heap:}
         We present a new algorithm for the computation of augmented merge
         trees. \Charles{Contrary to massively parallel
         approaches~\cite{maadasamy12,acharya15,carr16}, our strategy
         revisits
         the optimal sequential algorithm for augmented
         trees~\cite{carr00}.
         A major distinction with the latter algorithm is the localized nature
         of our approach, based on} local sorting traversals whose results are
         progressively merged with the help of a Fibonacci heap. In this
         context, we also introduce a new criterion for the detection of the
         saddles which generate branching in the output tree, as well as an
         efficient procedure to process the output arcs in the vicinity of the
         root of the tree (hereafter referred to as the \emph{trunk}). Our
         algorithm is simple to implement and it improves practical time
         performances over a reference implementation~\cite{libtourte} of the
         traditional algorithm~\cite{carr00}.
      }
   \item{\textbf{Parallel augmented merge trees:}
         We show how to leverage the task runtime environment of OpenMP to
         easily implement a shared-memory\Charles{, coarse-grained} parallel
         version of the above algorithm \Charles{for multi-core
           architectures}. Instead of introducing extra work with
         a static decomposition of the mesh among the threads, the local
         algorithm based on Fibonacci heaps naturally distributes the merge
         tree arc computations via independent tasks on the CPU cores. We hence
         avoid any extra work in parallel, while enabling an efficient dynamic
         load balancing on the CPU cores thanks to the task runtime. This
         results in superior time and scaling performances compared to previous
         multi-threaded algorithms for augmented merge
         trees~\cite{gueunet2016contour}.
      }
\end{enumerate}

\noindent
This extended version makes these additional contributions.
\begin{enumerate}
   \setcounter{enumi}{2}
   \item{\textbf{Complete taskification:} We express every parallel work
         for our entire approach using tasks and nested parallelism. This complete
         taskification enables us to overlap tasks arising from the concurrent
         computations of the join and split trees. In practice this allows the
         runtime to pick tasks from one of the two trees if the other is running
         out of work, thus improving the parallel efficiency.
      }
   \item{\textbf{Parallel combination of the join and split
         trees:} We present a new parallel algorithm to combine the join and
         split trees into the output contour tree. First, we describe a
         procedure to combine arcs in parallel \Charles{which exploits nested
         parallelism}. Second, to further speedup this step, we introduce a new
         original method  for the fast parallel processing of the arcs on the
         trunk of the tree. \Charles{Detailed performance results concerning
         this parallel combination are given and analyzed.}
      }
   \item {\textbf{Fine grain optimizations:} We provide several optimizations
         reducing the amount of work of our algorithm in practice. First, we
         show how to trigger  the efficient trunk computation earlier using an
         improved detection. Second, we show how to avoid the valence
         processing on most vertices with lazy evaluation at saddle points
         only. Finally, we document how switching form a structure of arrays
         (SoA) to an array of structures (AoS) contributes to performance
         improvement.
      }
   \item{\textbf{Implementation:}
         We provide a lightweight VTK-based C++ implementation of our approach
         for reproducibility purposes.
      }
\end{enumerate}

\section{Preliminaries}%
\label{section:Preliminaries}

This section briefly describes our formal setting and presents an overview of
our approach. An introduction to topological data analysis can be found
in~\cite{edelsbrunner09, tiernyBook18}.

\subsection{Background}%
\label{section:Background}

The input to our algorithm is a piecewise linear (PL) scalar field $f : \domain
\rightarrow \range$ defined on a simply-connected PL $d$-manifold~$\domain$
(i.e.\ a triangular mesh for $d=2$, or a tetrahedral one for $d=3$). This scalar
field generally corresponds to the results of a numerical simulation or of an
acquisition evaluated on each vertex. Without loss of generality, we will
assume that $d = 3$ (tetrahedral meshes) in most of our discussion, although
our algorithm supports arbitrary dimensions.
An \emph{$i$-simplex} of $\domain$ denotes a vertex ($i=0$), an edge ($i=1$), a
triangle ($i=2$) or a tetrahedron ($i=3$). Then, the \emph{star} $\st{v}$ of a
vertex $v$ is the set of simplices of $\domain$ which contain $v$.
The \emph{link} $\lk{v}$ is the set of faces (i.e.\ sub-simplices) of the
simplices of $\st{v}$ which do not intersect $v$. Intuitively, the link of a
vertex $v$ in 2D consists of the ring of edges immediately around $v$. In 3D,
it corresponds to the sphere of triangles immediately around it. We will note
$Lk_i(v)$ the set of $i$-simplices of $\lk{v}$. The scalar field $f$ is
provided on the vertices of $\domain$ and it is linearly interpolated on the
simplices of higher dimension. We will additionally require that the
restriction of $f$ to the vertices of $\domain$ is injective, which can be
easily enforced with a mechanism inspired by simulation of
simplicity~\cite{edelsbrunner90}.

\begin{figure}
    \centering
    \includegraphics[width=\linewidth]{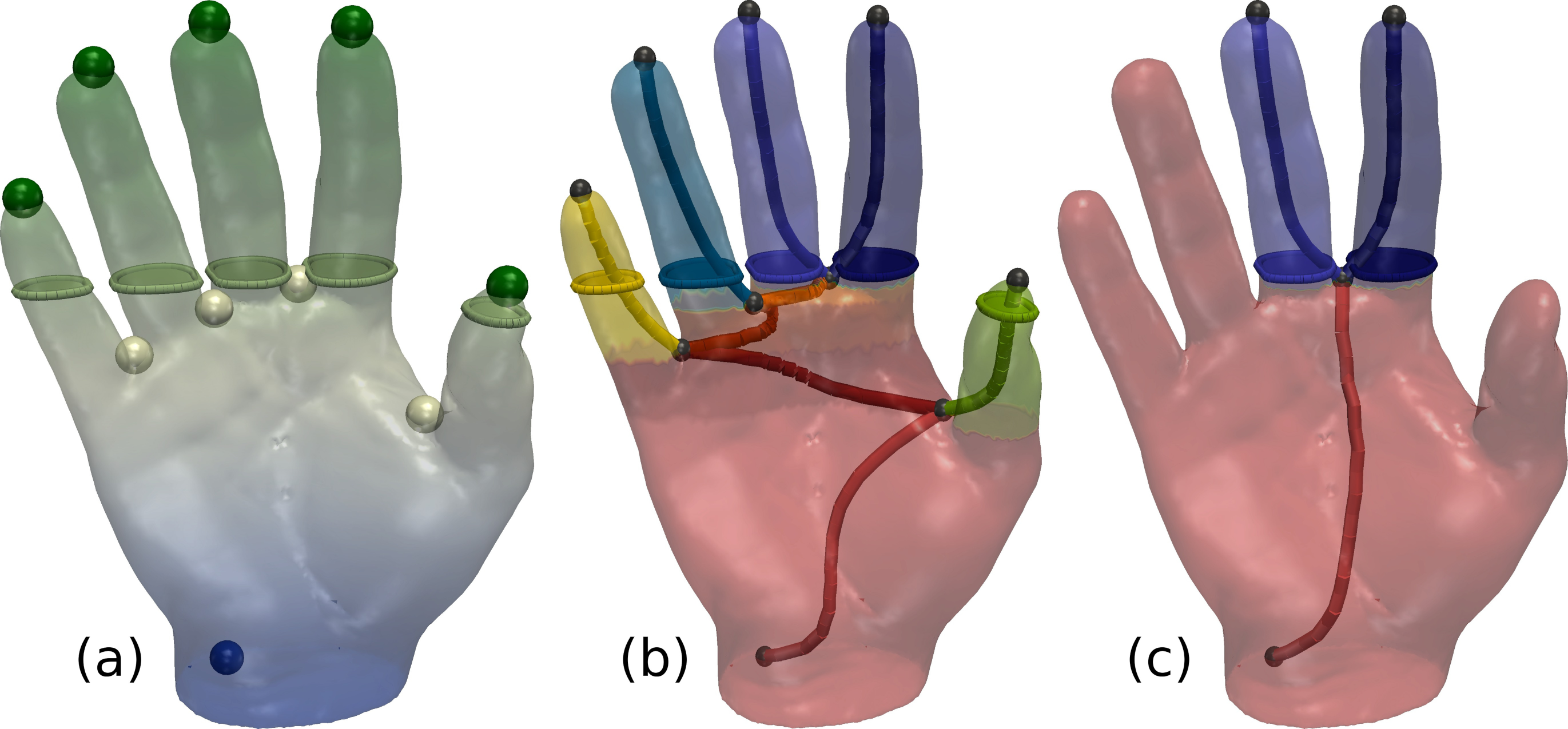}
    \vcaption{%
        Topology driven hierarchical data segmentation. (a) Input field $f$
        (color map), level-set (light green) and critical points (blue:
        minimum, white: saddle, green: maximum). (b) Split tree of $f$ and its
        corresponding segmentation (arcs and their pre-images by $\phi$ are
        shown with the same color). (c) Split tree of $f$ and its corresponding
        segmentation, simplified according to persistence.
    }\label{fig:background}
\end{figure}

The notion of critical point from the smooth setting~\cite{milnor63} admits a
direct counterpart for PL scalar fields~\cite{banchoff70}. Let $\lkminus{v}$ be
the \emph{lower link} of the vertex $v$: $\lkminus{v} = \{\sigma \in
\lk{v}~|~\forall u \in \sigma : f(u) < f(v)\}$. The \emph{upper link}
$\lkplus{v}$ is given by $\lkplus{v} = \{ \sigma \in \lk{v}~|~\forall u \in
\sigma : f(u) > f(v)\}$. Then, given a vertex $v$, if its lower (respectively
upper) link is empty, $v$ is a local \emph{minimum} (respectively
\emph{maximum}). If both $\lkminus{v}$ and $\lkplus{v}$ are simply connected,
$v$ is a regular point. Any other configuration is called a \emph{saddle}
(white spheres, \autoref{fig:background}a).

A level-set is defined as the pre-image of an isovalue $i \in \range$ onto
$\domain$ through $f$: $f^{-1}(i) = \{ p \in \domain~|~f(p) = i\}$
(\autoref{fig:background}a). Each connected component of a level-set is called
a \emph{contour}.  In \autoref{fig:background}b, each contour of
\autoref{fig:background}a is shown with a distinct color. Similarly, the notion
of \emph{sub-level set}, noted $\sub{i}$, is defined as the pre-image of the
open interval $(-\infty, i)$ onto $\domain$ through $f$: $\sub{i} = \{ p \in
\mathcal{M}~|~f(p) < i \}$. Symmetrically, the \emph{sur-level set} $\sur{i}$
is defined by $\sur{i} = \{ p \in \mathcal{M}~|~f(p) > i \}$. Let
$f^{-1}_{-\infty}\big{(f(p)\big)}_p$ (respectively
$f^{-1}_{+\infty}\big{(f(p)\big)}_p$) be the connected component of sub-level
set (respectively sur-level set) of $f(p)$ which contains the point $p$. The
\emph{split tree} $\splitTree$ is a 1-dimensional simplicial complex
(\autoref{fig:background}a) defined as the quotient space $\splitTree = \domain
/ \sim$ by the equivalence relation $p_1 \sim p_2$:
\begin{center}
    $
    \left\lbrace
    \begin{array}{l}
    f(p_1) = f(p_2)\\
    p_2 \in f^{-1}_{+\infty}\big{(f(p_1)\big)}_{p_1}
    \end{array}
    \right.
    $
\end{center}
Intuitively, the split tree is a tree which tracks the creation of connected
components of sur-level sets at its leaves and which tracks their merges at its
interior nodes (\autoref{fig:background}b). For regular isovalues (colored
surfaces in \autoref{fig:background}b), it contracts components to points on
the arcs connecting its nodes. The \emph{join tree}, noted $\joinTree$, is
defined similarly with regard to an equivalence relation on sub-level set
components (instead of sur-level sets), and tracks the merges of these
sub-level set connected components. Irrespective of their orientation, the
\emph{join} and \emph{split} trees are usually called \emph{merge trees}, and
noted $\mt$ in the following. The notion of \emph{Reeb graph}~\cite{reeb46},
noted $\reeb$, is also defined similarly, with regard to an equivalence
relation on level set components (instead of sub-level set components). As
discussed by Cole-McLaughlin et al.~\cite{cole03}, the construction of the Reeb
graph can lead to the removal of $1$-cycles, but not to the creation of new
ones.  This means that the Reeb graphs of PL scalar fields defined on
simply-connected domains are loop-free. Such a Reeb graph is called a
\emph{contour tree} and we will note it $\ct$. Contour trees can be computed
efficiently by combining the join and split trees with a linear-time
traversal~\cite{tarasov98, carr00}. In \autoref{fig:background}, since
$\domain$ is simply connected, the contour tree $\ct$ is also the Reeb graph of
$f$. Since $f$ has only one minimum, the split tree $\splitTree$ is equivalent
to the contour tree $\ct$.

Note that $f$ can be decomposed into $f = \psi~\circ~\phi$ where $\phi~:
\domain~\rightarrow~\mt$  maps each point in $\domain$ to its equivalence class
in $\mt$ and where $\psi~: \mt~\rightarrow~\range$ maps each point in $\mt$ to
its $f$ value. Since the number of connected components of $\sub{i}$, $\sur{i}$
and $f^{-1}(i)$ only changes in the vicinity of a critical
point~\cite{milnor63, banchoff70, edelsbrunner09}, the pre-image by $\phi$ of
any vertex of $\joinTree$, $\splitTree$ or $\reeb$ is a critical point of $f$
(spheres in \autoref{fig:background}a). The pre-image of vertices of valence 1
necessarily correspond to extrema of $f$~\cite{reeb46}. The pre-image of
vertices of higher valence correspond to saddle points which join (respectively
split) connected components of sub- (respectively sur-) level sets. Since
$\sub{f(M)} = \domain$ for the global maximum $M$ of $f$, $\phi(M)$ is called
the \emph{root} of $\joinTree$ and the image by $\phi$ of any local minimum $m$
is called a \emph{leaf}. Symmetrically, the global minimum of $f$ is the root
of $\splitTree$ and local maxima of $f$ are its leaves.

Note that the pre-image by $\phi$ of $\ct$ induces a partition of $\domain$.
The pre-image $\phi^{-1}( \sigma_1 )$  of an arc $ \sigma_1  \in \ct$ is
guaranteed by construction to be connected.  This latter property is at the
basis of the usage of the contour tree in visualization as a data segmentation
tool (\autoref{fig:background}b) for feature extraction. In practice,
$\phi^{-1}$ is represented explicitly by maintaining, for each arc $ \sigma_1
\in \ct$, the list of regular vertices of $\domain$ that map to $\sigma_1 $.
Moreover, since the contour tree is a simplicial complex, persistent homology
concepts \cite{edelsbrunner02} can be readily applied to it by considering a
filtration based on $\psi$.  Intuitively,  this progressively simplifies $\ct$,
by iteratively removing its \emph{shortest} arcs connected to leaves. This
yields hierarchies of contour trees that are accompanied by hierarchies of data
segmentations, that the user can interactively explore in practice (see
\autoref{fig:background}c).

\Charles{In the following, we briefly describe two data structures used in the
core of our algorithm.  First, a Union-Find~\cite{cormen09} is a
data structure implementing
two operations
(\emph{union} and \emph{find}) and operating on disjoint sets to track
whether some elements are in the same
connected component
or not. Internally, it
relies on rooted trees and uses path compression along with a ranking
mechanism for improved efficiency, leading to $O\big(\alpha(n)\big)$ time
complexity per operation, where $\alpha()$ is the extremely slow-growing
inverse of the Ackermann function.}
\Charles{Second, the Fibonacci heap data structure, extensively used in our new
approach, is a priority queue introduced by M. Fredman and R.
Tarjan~\cite{fredman87,cormen09}. It is based on a collection of (binomial)
trees and offers constant time operations thanks to an advanced laziness
mechanism (in particular for the merge of two heaps), except for the \emph{pop}
operation which takes $O\big(log(n)\big)$ steps.}

\subsection{Overview}
\label{ssection:Overview}

\begin{figure*}
  \includegraphics[width=\linewidth]{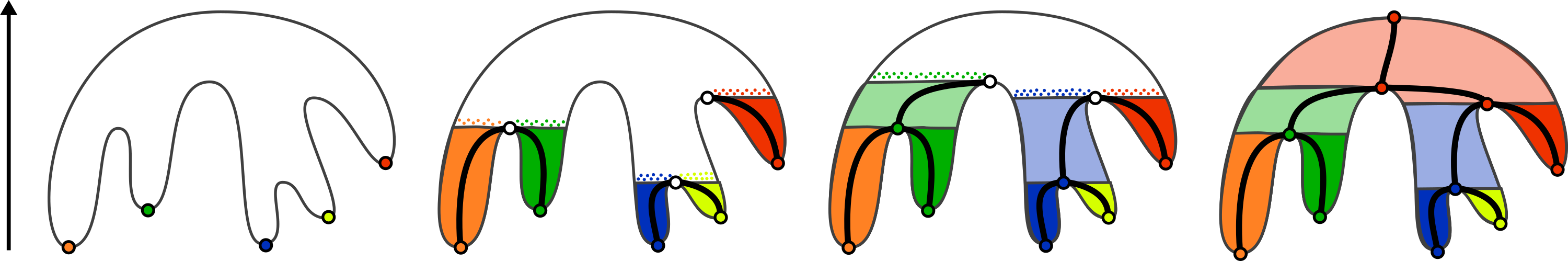}
  \vcaption{%
      Overview of our augmented merge tree algorithm based Fibonacci heaps (2D
      toy elevation example). First, the local extrema of $f$ (corresponding to
      the leaves of $\mt$) are extracted (left, \autoref{section:leafSearch}).
      Second, the arc $\sigma_m$ of each extremum $m$ is grown independently
      along with its segmentation (matching colors, center left,
      \autoref{section:leafGrowth}). These independent growths are achieved by
      progressively growing the connected components of level sets created in
      $m$, for increasing $f$ values, and by maintaining at each step a
      priority queue $\priorityQueue_m$, implemented with a Fibonacci heap,
      which stores vertex candidates for the next iteration (illustrated with
      colored dots). These growths are stopped at merge saddles (white disks,
      center left, \autoref{section:saddleStop}). Only the last growth reaching
      a saddle $s$ is kept active and allowed to continue to grow the saddle's
      arc $\sigma_s$ (matching colors, center right,
      \autoref{section:saddleGrowth}). The constant time merge operation of the
      Fibonacci heap (to initialize the growth at $s$) enables a highly
      efficient execution for this step in practice. Last, when only one growth
      remains active, the tree is completed by simply creating its
      \emph{trunk}, a monotone sequence of arcs to the root of the tree which
      links the remaining pending saddles (pale blue region, right,
      \autoref{section:trunkGrowth}). The task-based parallel model allows for
      a straightforward parallelization of this algorithm, where each arc is
      grown independently, only requiring local synchronizations on merge
      saddles.
  }
  \label{fig:overview}
\end{figure*}

An overview of our augmented merge tree computation algorithm is presented in
\autoref{fig:overview} in the case of the join tree. The purpose of our
algorithm, in addition to construct $\mt$, is to build the explicit
segmentation map $\phi$, which maps each vertex $v \in~\domain$ to $\mt$. Our
algorithm is expressed as a sequence of procedures, called on each vertex of
$\domain$. First, given a vertex $v$, the algorithm checks if $v$ corresponds
to a leaf (\autoref{fig:overview} left, \autoref{section:leafSearch}). If this
is the case, the second procedure is triggered. For each leaf vertex, the
augmented arc connected to it is constructed by a local growth, implemented
with a sorted breadth-first search traversal (\autoref{fig:overview} middle
left, \autoref{section:leafGrowth}). A local growth may continue at a join
saddle $s$, in a third procedure, only if it is the last growth which visited
the saddle $s$ (\autoref{fig:overview} middle right,
\autoref{section:saddleGrowth}). To initiate the growth from $s$ efficiently,
we rely on the Fibonacci heap data-structure~\cite{fredman87, cormen09} in our
breadth-first search traversal, which supports constant-time merges of sets of
visit candidates. A fourth procedure (the trunk growth) is triggered to
abbreviate the process when a local growth happens to be the last active
growth. In this case, all the unvisited vertices above $s$ are guaranteed to
map through $\phi$ to a monotone path from $s$ to the root
(\autoref{fig:overview} right, \autoref{section:trunkGrowth}). Overall, the
time complexity of our algorithm is identical to that of the reference
algorithm \cite{carr00}: $O(\vertexNumber~\log(\vertexNumber) + \edgeNumber
\alpha(\edgeNumber)\big)$, where $|\sigma_i|$ stands for the number of
$i$-simplices in $\domain$ and $\alpha()$ is the inverse of the Ackermann
function (\Charles{cf. \autoref{section:Background}}).

For the augmented contour tree computation, we present a new parallel
combination algorithm which improves the sequential reference
method~\cite{carr00}. Our algorithm processes the arcs from the join and split
trees in parallel, level by level. Once the join and split trees only
contribute one arc each, the remaining work only consists in completing the
output tree with a set of arcs forming a monotone path. We use the fourth
procedure of our merge tree algorithm to process in parallel these remaining
arcs and vertices.

\section{Merge trees with Fibonacci heaps}
\label{section:Algorithm}

In this section, we present our algorithm for the computation of augmented
merge trees based on local arc growth. Our algorithm consists of a sequence of
procedures applied to each vertex, described in each of the following
subsections. In the remainder, we illustrate our discussion with the join tree,
which tracks connected components of sub-level sets, initiated in local minima.

\subsection{Leaf search}
\label{section:leafSearch}

\begin{algorithm}
   \caption{\Charles{Find minima of the input mesh}}
   \label{algo:leafSearch}
   \CharlesV{%
   \begin{algorithmic}
         \Procedure{LeafSearch}{Mesh: $\domain$}
         \For{each vertex $v \in~\domain$}
            \Comment{in parallel (tasks)}
            \State{add $v$ to leaves if $|\lkminusV{v}| = 0$}
         \EndFor
         \State{\textbf{return} leaves}
         \EndProcedure
   \end{algorithmic}
   }
\end{algorithm}

\Charles{The procedure \textbf{LeafSearch} is used to find the minima on which
local growths will later be initiated is shown in
Alg.~\autoref{algo:leafSearch}.}
\Charles{Minima are vertices with an empty lower link: $|\lkminusV{v}| = 0$.}

\subsection{Leaf growth}
\label{section:leafGrowth}

\begin{algorithm}
   \caption{\Charles{Local growth computing one arc of $\mt$}}
   \label{algo:arcGrowth}
   \CharlesV{%
   \begin{algorithmic}
      \Procedure{ArcGrowth}{$\priorityQueue_m$: Fibonacci heap, uf: Union-Find}
      \State{Open a new arc in $\mt$ at the first vertex of $\priorityQueue_m$}
      \While{not the last active growth}
      \State{Pop the first vertex of $\priorityQueue_m$ in $v$}
         \State{Process $v$}
         \State{Add $\lkplusV{v}$ into the $\priorityQueue_m$}
         \State{Use $\lkminusV{v}$ to check if $v$ is a merging saddle}
         \If{$v$ is a merging saddle}
            \If{last growth reaching $v$} 
               \State{SaddleGrowth($v$)}
            \EndIf
            \State{\textbf{return}}
         \EndIf
      \EndWhile
      \EndProcedure
   \end{algorithmic}
   }
\end{algorithm}

\Charles{For each} local minimum $m$, the \Charles{leaf} arc $\sigma_m$ of the join tree
connected to it is constructed with a procedure that we call
\Charles{\textbf{ArcGrowth},
presented in Alg.~\autoref{algo:arcGrowth}}. The purpose of this procedure
is to progressively sweep all contiguous equivalence classes
(\autoref{section:Background}) from $m$ to the saddle $s$ located at the
extremity of $\sigma_m$. We describe how to detect such a saddle $s$, and
therefore where to stop such a growth, in the next subsection
(\autoref{section:saddleStop}). In other words, this growth procedure will
construct the connected component of sub-level set initiated in $m$, and will
make it progressively grow for increasing values of $f$.

This is achieved by implementing an ordered breadth-first search traversal of
the vertices of $\domain$ initiated in $m$. At each step, the neighbors of $v$
which have not already been visited are added to a priority queue
$\priorityQueue_m$ (if not already present in it), \Charles{implemented as a
Fibonacci heap~\cite{fredman87,cormen09}. Additionally, $v$ is
\textit{processed} by the current growth: the vertex is marked with the
identifier of the current arc $\sigma_m$ for future addition.} The purpose of
the addition of $v$ to $\sigma_m$ is to augment this arc with regular vertices,
and therefore to store its data segmentation. Next, the following visited
vertex $v'$ is chosen as the minimizer of $f$ in $\priorityQueue_m$ and the
process is iterated until $s$ is reached (\autoref{section:saddleStop}). At
each step of this local growth, since breadth-first search traversals grow
connected components, we have the guarantee, when visiting a vertex $v$, that
the set of vertices visited up to this point (added to $\sigma_m$) indeed
equals to the set of vertices belonging to the connected component of sub-level
set of $f(v)$ which contains $v$, noted $f^{-1}_{-\infty}\big(f(v)\big)_{v}$ in
\autoref{section:Background}. Therefore, our local leaf growth indeed
constructs $\sigma_m$ (with its segmentation). Also, note that, at each
iteration, the set of edges linking the vertices already visited and the
vertices currently in the priority queue $\priorityQueue_m$ are all crossed by
the level set  $f^{-1}\big(f(v)\big)$.

The Time complexity of this procedure is $O(\vertexNumber~log(\vertexNumber) +
\edgeNumber)$, where $|\sigma_i|$ stands for the number of $i$-simplices in
$\domain$.

\subsection{Saddle stopping condition}
\label{section:saddleStop}

Given a local minimum $m$, the leaf growth procedure is stopped when reaching
the saddle $s$ corresponding to the other extremity of $\sigma_m$. We describe
in this subsection how to detect $s$.

In principle, the saddles of $f$ could be extracted by using the critical point
extraction procedure presented in \autoref{section:Background}, based on a
local classification of the link of each vertex. However, such a strategy has
two disadvantages. First not all saddles of $f$ necessarily corresponding to
branching in $\joinTree$ and/or $\splitTree$. Thus some unnecessary computation
would need to be carried out. Second, we found in practice that even optimized
implementations of such a classification~\cite{ttk} tend to be slower than the
entire augmented merge tree computation in sequential. Thus, another strategy
should be considered for the sake of performance.

\begin{figure}
  \includegraphics[width=\linewidth]{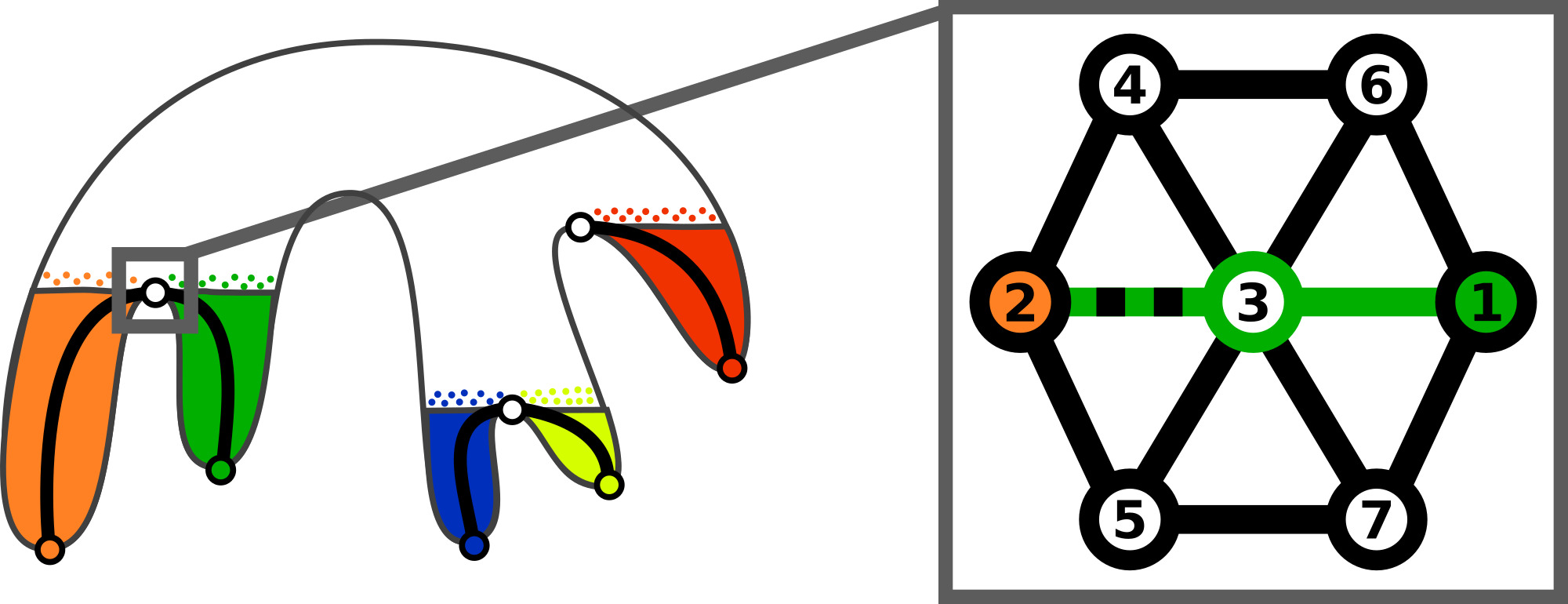}
  \vcaption{%
      Local merge saddle detection based on arc growth (2D elevation example
      from \autoref{fig:overview}). The local growth of the arc $\sigma_{m}$
      (green) will visit the vertex $v'$ at value 3 after visiting the vertex
      at value 1 (following the priority queue $\priorityQueue_m$). At this
      point, the neighbors of $v'$ which have not been visited yet by
      $\sigma_{m}$ and which are not in $\priorityQueue_{m}$ yet (dashed green
      edges) will be added to $\priorityQueue_m$. The minimizer $v$ of
      $\priorityQueue_m$ (vertex 2) has a scalar value lower than $v'$. Hence
      $v'$ is a merge saddle.
   }
   \label{fig:saddleDetection}
\end{figure}

The local \Charles{\textbf{ArcGrowth}} procedure (\autoref{section:leafGrowth})
visits the vertices of $\domain$ with a breadth-first search traversal
initiated in $m$, for increasing $f$ values. At each step, the minimizer $v$ of
$\priorityQueue_m$ is selected. Assume that \Charles{at some point:} $f(v) <
f(v')$ where $v'$ was the vertex visited immediately before $v$. This implies
that $v$ belongs to the lower link of $v'$, $\lkminus{v'}$. Since $v$ was
visited after $v'$, this means that $v$ does not project to $\sigma_m$ through
$\phi$. In other words, this implies that $v$ does not belong to the connected
component of sub-level set containing $m$. Therefore, $v'$ happens to be the
saddle $s$ that correspond to the extremity of $\sigma_m$. Locally
(\autoref{fig:saddleDetection}), the local leaf growth entered the star of $v'$
through the connected component of lower link projecting to $\sigma_m$ and
jumped across the saddle $v'$ downwards when selecting the vertex $v$, which
belongs to another connected component of lower link of $v'$.

Therefore, a sufficient condition to stop \Charles{an arc growth}
is
when the candidate vertex returned by the priority queue has a lower $f$ value
than the vertex visited last. In such a case, the last visited vertex is the
saddle $s$ which closes the arc $\sigma_m$ (\autoref{fig:saddleDetection}).

\subsection{Saddle growth}
\label{section:saddleGrowth}

\begin{algorithm}
  \caption{\Charles{Start a local growth at a join saddle}}
   \label{algo:saddleGrowth}
   \CharlesV{%
   \begin{algorithmic}
      \Procedure{SaddleGrowth}{$s$: join saddle}
         \State{Close arcs in $\lkminus{s}$}
         \State{$\priorityQueue_m \leftarrow$ union $\priorityQueue_{m_0}, \priorityQueue_{m_1},\ldots \priorityQueue_{m_n} \in \lkminusV{s}$}
         \State{uf~~~~$\leftarrow$ union uf$_{0}$, uf$_{1}$,\ldots uf$_{n} \in \lkminusV{s}$}
         \State{ArcGrowth($\priorityQueue_m$, uf)}
      \EndProcedure
   \end{algorithmic}
   }
\end{algorithm}

Up to this point, we described how to construct each arc $\sigma_m$ connected
to a local minimum $m$, along with its corresponding data segmentation. The
remaining arcs can be constructed similarly.

\begin{figure}
  \includegraphics[width=\linewidth]{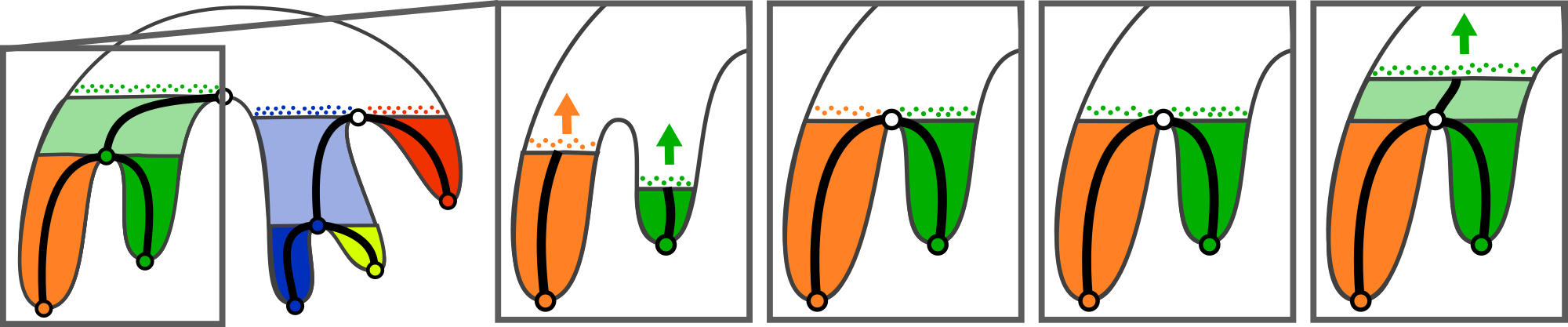}
  \vcaption{%
      Union of priority queues at a merge saddle (2D elevation example from
      \autoref{fig:overview}). Initially, each arc growth maintains its own
      priority queue (illustrated with colored dots, left inset). When reaching
      a merge saddle $s$ (second inset), the growths which arrived first in $s$
      are marked \emph{terminated}. Only the last one (green) will be allowed
      to resume the growth from $s$ to construct the arc $\sigma_s$ (last
      inset). To continue the propagation of the sub-level set component which
      contains $s$, the priority queues of all growths arrived at $s$ need to
      be merged into only one (third inset) prior to resuming the propagation.
      If done naively, this operation could yield a quadratic runtime
      complexity for our approach overall. Since Fibonacci heaps support
      constant time merges, they guarantee the linearithmic complexity of our
      overall approach.
  }
  \label{fig:frontMerge}
\end{figure}

Given a local minimum $m$, its leaf growth is stopped at the saddle $s$ which
corresponds to the extremity of the arc connected to it, $\sigma_m$. When
reaching $s$, if \emph{all} vertices of $\lkminus{s}$ have already been visited
by some local leaf growth, we say that the current growth, initiated in $m$, is
the \emph{last} one visiting $s$. In such a case, the \Charles{procedure
\textbf{SaddleGrowth} presented in Alg.~\autoref{algo:saddleGrowth} is
called (see Algorithm \autoref{algo:arcGrowth}) and the} same
  breadth-first search traversal can be applied to grow the
arc of $\joinTree$ initiated in $s$, noted $\sigma_s$. In order to represent
all the connected components of sub-level set merging in $s$, such a traversal
needs to be initiated with the \emph{union} of the priority queues
$\priorityQueue_{m_0}, \priorityQueue_{m_1}, \dots \priorityQueue_{m_n}$ of
\emph{all} the arcs merging in $s$. Such a union models the entire set of
candidate vertices for absorption in the sub-level component of $s$
(\autoref{fig:frontMerge}). Since both the number of minima of $f$ and the size
of each priority queue can be linear with the number of vertices in $\domain$,
if done naively, the union of all priority queues could require
$O(\vertexNumber^2)$ operations overall. To address this issue, we model each
priority queue with a Fibonacci heap~\cite{fredman87, cormen09}, which supports
the removal of the minimizer of $f$ from $\priorityQueue_m$ in
$log(\vertexNumber)$ steps, and performs both the insertion of a new vertex and
the merge of two queues in constant time.

Similarly to the traditional merge tree algorithm~\cite{tarasov98, carr00}, we
maintain a Union-Find data structure to precisely keep track of the arcs which
need to be merged at a given saddle $s$. Each local minimum $m$ is associated
with a unique Union-Find element, which is also associated to all regular
vertices mapped to $\sigma_m$ (\autoref{section:leafGrowth}). Also, each
Union-Find element is associated to the arc it currently grows. When an arc
$\sigma$ reaches a join saddle $s$ last, the find operation of the Union-Find
is called on each vertex of $\lkminus{s}$ to retrieve the set of arcs which
merge there, and the union operation is called on \Charles{all Union-Find
associated to these arcs} to keep track of the merge event. Thus,
overall, the time complexity of our augmented merge tree computation is
$O\big(\vertexNumber log(\vertexNumber) + \edgeNumber \alpha(\edgeNumber)\big)$.
The $\edgeNumber \alpha(\edgeNumber)$ term yields
from the usage of the Union-Find data structure, while the Fibonacci heap,
thanks to its constant time merge support, enables to grow the arcs of the tree
in logarithmic time. The time complexity of our algorithm is then exactly
equivalent to the traditional algorithm~\cite{tarasov98, carr00}. However,
comparisons to a reference implementation~\cite{libtourte}
(\autoref{section:Results}) show that our approach provides superior
performances in practice.

\subsection{Trunk growth}
\label{section:trunkGrowth}

\begin{algorithm}
   \caption{\Charles{Compute the last monotone path}}
   \label{algo:trunk}
   \CharlesV{%
   \begin{algorithmic}
      \Procedure{Trunk}{}
      \State{Close arcs on pending saddles}
      \State{Create a monotone path from the last visited vertex to the global maximum}
      \For{each unvisited vertex $v_u$}
         \Comment{in parallel (tasks)}
         \State{Project $v_u$ into its arc on the monotone path}
      \EndFor
      \EndProcedure
   \end{algorithmic}
   }
\end{algorithm}

Time performance can be further improved by abbreviating the process when only
one arc growth is remaining. Initially, if $f$ admits $N$ local minima, $N$
arcs (and $N$ arc growths) need to be created. When the growth of an arc
$\sigma$ reaches a saddle $s$, if $\sigma$ is not the last arc reaching $s$,
the growth of $\sigma$ is switched to the \emph{terminated} state. Thus, the
number of remaining arc growths will decrease from $N$ to $1$ along the
execution of the algorithm. In particular, the last arc growth will visit all
the remaining, unvisited, vertices of $\domain$ upwards until the global
maximum of $f$ is reached, possibly reaching on the way an arbitrary number of
\emph{pending} join saddles, where other arc growths have been stopped and
marked terminated (white disks, \autoref{fig:overview}, third column). Thus,
when an arc growth reaches a saddle $s$, if it is the last active one, we have
the guarantee that it will construct in the remaining steps of the algorithm a
sequence of arcs which constitutes a monotone path from $s$ up to the root of
$\joinTree$. We call this sequence the \emph{trunk} of $\joinTree$
(\autoref{fig:overview}) \Charles{and we present the corresponding procedure in
Alg.~\autoref{algo:trunk}}. The trunk of the join tree can be computed
faster than through the breadth-first search traversals described in
Secs.~\ref{section:leafGrowth} and~\ref{section:saddleGrowth}. Let $s$ be the
join saddle where the trunk starts. Let $S = \{s_0, s_1, \dots s_n\}$ be the
sorted set of join saddles that are still pending in the computation (which
still have unvisited vertices in their lower link). The trunk is constructed by
simply creating arcs that connect two consecutive entries in $S$. Next, these
arcs are augmented by simply traversing the vertices of $\domain$ with higher
scalar value than $f(s)$ and projecting each unvisited vertex $v_u$ to the trunk
arc that spans it scalar value $f(v_u)$.

Thus, our algorithm for the construction of the trunk does not use any
breadth-first search traversal, as it does not depend on any mesh traversal
operation, and it is performed in $O(\vertexNumber log(\vertexNumber))$ steps
(to maintain regular vertices sorted along the arcs of the trunk).
This algorithmic step is another important novelty of our approach.

\Charles{Finally, the overall merge tree computation is presented in
Alg.~\autoref{algo:mt}}

\begin{algorithm}
   \caption{\Charles{Overall merge tree computation for a mesh $\domain$}}
   \label{algo:mt}
   \CharlesV{%
   \begin{algorithmic}
      \State{leaves $\leftarrow$ LeafSearch($\domain$)}
      \For{each $v \in$ leaves}
         \State{$\priorityQueue_m \leftarrow$ new Fibonacci heap containing $v$}
         \State{uf~~~~$\leftarrow$ new Union-Find}
         \State{ArcGrowth($\priorityQueue_m$, uf)}
         \Comment{task} 
      \EndFor
      \State{Trunk()}
   \end{algorithmic}
   }
\end{algorithm}

\section{Task-based parallel merge trees}
\label{section:Parallel}

The previous section introduced a new algorithm based on local arc growths with
Fibonacci heaps for the construction of augmented join trees (split trees being
constructed with a symmetric procedure). Note that this algorithm enables to
process the minima of $f$ concurrently. The same remark goes for the join
saddles; however, a join saddle growth can only be started after all of its
lower link vertices have been visited. Such an independence and synchronization
among the numerous arc growths can be straightforwardly parallelized thanks to
the task parallel programming paradigm. Also, note that such a split of the
work load does not introduce any supplementary computation \Charles{or memory
overhead}. Task-based runtime environments also naturally support dynamic load
balancing, each available thread picking its next task among the unprocessed
ones. We rely here on OpenMP tasks~\cite{openmp4.5}, but other task runtimes
(e.g. Intel Threading Building Blocks, Intel Cilk Plus, etc.) could be used as
well with a few modifications. In practice, users only need to specify a number
of threads among which the tasks will be scheduled. In the remainder, we will
\Charles{present our taskification process for the merge tree computation, as
well as the required task synchronizations.}

At a technical level, our implementation starts with a global sort of all the
vertices according to their scalar value in parallel (using the GNU parallel
sort). This reduces further vertex comparisons to comparisons of indices, which
is faster in practice than accessing the actual scalar values and which is also
independent of the scalar data representation. Our experiments have shown that
this sort benefits from a better data locality, and  is thus more efficient,
when using an array of structures (AoS) rather than a structure of arrays (SoA)
for the vertex data structures (id, scalar value, offset). The remaining steps
of our approach being unsuitable for SIMD computing and mostly consisting of
scattered memory accesses, the shift to the AoS data layout did not affect
their performance.

\subsection{\Charles{Taskification}}

\paragraph*{\textbf{Parallel leaf search}} \label{section:paraLeafSearch} For
each vertex $v \in~\domain$, the extraction of its lower link $\lkminus{v}$ is
makes this step embarrassingly parallel and enables a straightforward
parallelization of the corresponding loop using OpenMP tasks\Charles{:
  see Alg.~\autoref{algo:leafSearch}}. Once done, we
have the list of extrema from which the leaf growth should be started. This
list is sorted so that the leaf growths are launched in the order of the scalar
value of their extremum, starting with the ``deepest'' leaves.

\paragraph*{\textbf{\Charles{Arc growth tasks}}} Each \Charles{arc} growth is
independent from the others, spreading locally until it finds a saddle. Each
leaf growth is thus simply implemented as a task, starting at its previously
extracted leaf \Charles{as shown in Alg.~\autoref{algo:arcGrowth}}.
\Charles{All tasks but the last one stop at the next saddle: this last task then
proceeds with this saddle growth. }

\subsection{\Charles{Synchronizations}}

\Charles{In the following, we present the task synchronizations
  required for a parallel execution of our algorithm.}

\paragraph*{\textbf{Saddle stopping condition}} \label{section:paraSaddleStop}
The saddle stopping condition presented in \autoref{section:saddleStop} can be
safely implemented in parallel with tasks. When a vertex $v$, unvisited so far
by the current arc growth,  is visited immediately after a vertex $v'$ with
$f(v) < f(v')$, then $v'$ is a saddle. To decide if $v$ was indeed not visited
by an arc growth associated to the sub-tree of the current arc growth, we use
\Charles{the} Union-Find data structure \Charles{described
in \autoref{section:saddleGrowth}} (one Union-Find node per leaf). In
particular,
we store for each visited vertex the Union-Find representative of its current
growth (which was originally created on a minimum). Our Union-Find
implementation supports concurrent \emph{find} operations from parallel arc
growths (executed simultaneously by distinct tasks). A \emph{find} operation on
a Union-Find currently involved in a \emph{union} operation is also possible
but safely handled in parallel in our implementation. Since the \emph{find} and
\emph{union} operations are local to each Union-Find sub-tree~\cite{cormen09},
these operations generate only few concurrent accesses. Moreover, these
concurrent accesses are efficiently handled since only atomic operations are
involved.


\paragraph*{\textbf{\Charles{
  Detection of the last growth reaching a saddle}}} When a saddle $s$ is
detected, we also have to check if the current growth is the last to reach $s$
as described in \autoref{section:saddleGrowth}. For this, we rely on the size
of $\lkminusV{s}$, noted $|\lkminusV{s}|$ (number of vertices in the lower link
of $s$). In our preliminary approach~\cite{gueunet2017ftm}, this size was
computed for every vertices during the leaf search to avoid synchronization
issues. In contrast, the current approach strictly restrict this computation to
vertices where it is necessary and we address synchronization issues as
follows. Initially, a \emph{lower link counter} associated with $s$ is set to
$-1$. Each task $t$ reaching $s$ will atomically decrement this counter by
$n_t$, the number of vertices in $\lkminus{s}$ visited by $t$. Using here an
OpenMP atomic capture operation, only the first task reaching $s$ will retrieve
$-1$ as the initial value of $s$ (before the decrement). This  first task will
then compute $|\lkminusV{s}|$ and will (atomically) increment the counter by
$|\lkminusV{s}|+1$. Since the sum over $n_t$ for all tasks reaching $s$ equals
$|\lkminusV{s}|$, the task eventually setting the counter to 0 will be
considered as the ``last'' one reaching $s$ (note that it can also be the one
which computed $|\lkminusV{s}|$). We thus rely here only on lightweight
synchronizations, and avoid using a critical section.



\paragraph*{\textbf{\Charles{
      Growth merging at a saddle}}}
\label{sec:saddleGrowthSec}
Once the lower link of a saddle has been completely visited, the ``last'' task
which reached it merges the priority queues (implemented as Fibonacci heaps),
and the corresponding Union-Find data structures, of all tasks
\emph{terminated} at this saddle. Such an operation is  performed sequentially
at each saddle, without any concurrency issue both for the merge of the
Fibonacci heaps and for the \emph{union} operations on the Union-Find. The
saddle growth starting from this saddle is performed by this last task, with no
new task creation. This continuation of tasks is illustrated with shades of the
same color in \autoref{fig:overview} (in particular for the green and blue
tasks). As the number of tasks can only decrease, the detection of the trunk
start is straightforward. Each time a task terminates at a saddle, it
decrements atomically an integer counter, which tracks the number of remaining
tasks. The trunk starts when this number reaches one.

\paragraph*{\textbf{\Charles{Early trunk detection}}}
An early trunk detection procedure can be considered, in order for the last
active task to realize earlier, before reaching its upward saddle, that it is
indeed the last active task and therefore to trigger the efficient (and
parallel) trunk processing procedure even earlier. This detection consists in
regularly checking, within each local growth, if a task is the last active one
or not. In practice, we check the number of remaining tasks every 10,000
vertices on our experimental setup to avoid slowing down significantly the
computation. This improvement is particularly beneficial on data sets composed
of large arcs. In this case, a significant section of the arc previously
processed  by only one active task is now efficiently processed in parallel
during the trunk growth procedure.

\subsection{Parallel trunk growth}
\label{section:parallelTrunk}

During the arc growth step, we keep track of the \emph{pending} saddles
(saddles reached by some tasks but for which the lower link has not been
completely visited yet). The list of pending saddles enables us to compute the
trunk \Charles{in parallel 
  as described in Alg.~\autoref{algo:trunk}}.
Once the trunk growth has started, we only focus on the vertices whose scalar
value is strictly greater than the lowest pending saddle, as all other vertices
have already been processed during the regular arc growth procedure. Next, we
create the sequence of arcs connecting pairs of pending saddles in ascending
order. At this point, each vertex can be projected independently of the others
along one of these arcs. Using the sorted nature of the list of pending
saddles, we can use dichotomy for a fast projection. Moreover when we process
vertices in the sorted order of their index, a vertex can use the arc of the
previous one as a lower bound for its own projection: we just have to check if
the current vertex still projects in this arc or in an arc with a higher scalar
value. We parallelize this vertex projection procedure using tasks: each task
processes chunks of contiguous vertex indices out of the globally sorted vertex
list (see e.g. the OpenMP taskloop construct~\cite{openmp4.5}). For each chunk,
the first vertex is projected on the corresponding arc of the trunk using
dichotomy. Each new vertex processed next relies on its predecessor for its own
projection. Note that this procedure can visit (and ignore) vertices already
processed by the arc growth step.

\section{Task-based parallel contour trees}
\label{section:taskContourTree}

\begin{algorithm}
   \caption{\Charles{Overall contour tree computation for a mesh $\domain$}}
   \label{algo:ct}
   \CharlesV{%
   \begin{algorithmic}
      \State{LeafSearch($\domain$)}
      \State{
      \hspace{-0.3cm}$
      \left. 
      \begin{array}{l}
      \textnormal{Compute JT}\\
      \textnormal{Compute ST}
      \end{array}
      \right\rbrace
      $
      }
      \Comment{using 2 concurrent tasks\hspace{1.5cm}~}
      \State{Post-processing of the two merge trees}
      \State{ArcsCombine()}
      \State{TrunkCombine()}
   \end{algorithmic}
   }
\end{algorithm}

As described in \autoref{section:RelatedWork}, an important use case for the
merge tree is the computation of the contour tree. Our task-based merge tree
algorithm can be used quite directly for this purpose. First, \Charles{as shown
in Alg.~\autoref{algo:ct}} a single leaf search can be used to extract
both minima for the join tree and maxima for the split tree in a single
traversal instead of having each tree performing this step, thus avoiding one
pass on the data as done in our preliminary work~\cite{gueunet2017ftm}. Once
the two merge trees are computed (\autoref{section:Parallel}) while taking
here advantage of their concurrent processing, a post-processing step,
explained below, is required. Then, the two trees can be combined efficiently
into the contour tree using a new parallel combination algorithm.

\subsection{Post-processing for contour tree augmentation}
\label{section:postProcess}

Our merge tree procedure segments $\domain$ by marking each vertex with the
identifier of the arc it projects to through $\phi$. In order to produce such a
segmentation for the output contour tree (\autoref{section:paraCombination}),
each arc of $\mt$ needs to be equipped at this point with the explicit sorted
list of vertices which project to it. We reconstruct these explicit sorted
lists in parallel. For vertices processed by the arc growth step, we save
during each arc growth the visit order local to this growth. During the
parallel post-processing of all these vertices, we can safely build (with a
linear operation count) the ordered list of regular vertices of each arc in
parallel thanks to this local ordering.
Regarding the vertices
processed by the trunk step, we cannot rely on such a local ordering of the
arc. Instead each thread concatenates these vertices within bundles (one bundle
per arc for each thread). The bundles of a given arc are then sorted according
to their first vertex and concatenated in order to obtain the ordered list of
regular vertices for this arc. Hence, the $O(n \log{n})$ operation count of the
sort only applies to the number of bundles, which is much lower than the number
of vertices in practice. At this point, to use the combination pass the join
tree needs to be augmented with the nodes of the split tree and vice-versa.
This step is straightforward since each vertex stores the identifier of the arc
it maps to, for both trees. This short step can be done in parallel, using one
task for each tree.

\subsection{Tasks overlapping for merge tree computation}
\label{section:treeMix}

As discussed in \autoref{section:trunkGrowth}, during the arc growth step, the
number of  active tasks decreases monotonically and is driven by the topology
of the tree. When the number of remaining tasks to process becomes smaller than
the number of available threads, the computation enters a \textit{suboptimal
section}, where the parallel efficiency of our algorithm is undermined as some
threads are idle. During the contour tree computation the two merge trees are
computed using our task-based algorithm (\autoref{section:Parallel}). Contrary
to \cite{gueunet2017ftm}, we perform here a complete taskification of our
implementation, by always relying on tasks even when not required (see e.g. the
loop parallelization of the trunk growth in \autoref{section:parallelTrunk}).
It can be noticed that, in order to mitigate the cost of creating and managing
the tasks, a task is created in the different steps of the algorithm only when
the computation grain size is large enough (according to empirical thresholds).
As an example, each task is given a chunk of 400,000 vertices in the parallel
leaf search (\autoref{section:paraLeafSearch}).

This complete taskification enables us to lower the performance impact of the
suboptimal sections by computing the join and split trees
concurrently\Charles{: see Alg.~\autoref{algo:ct}}.
Indeed, when the computation of one of the two merge trees enters a suboptimal
section, the runtime can pick tasks from the other tree computation (from its
arc growth step, or from subsequent steps). By overlapping the two merge tree
computations, we can thus rely on more tasks to exploit at best the available
CPU cores.
In order to introduce such task overlap only when required, and thus to benefit
from it as long as possible, we also impose a higher priority on all tasks from
one of the two trees.



\subsection{Parallel Combination}
\label{section:paraCombination}

\begin{figure}[h]
  \includegraphics[width=\linewidth]{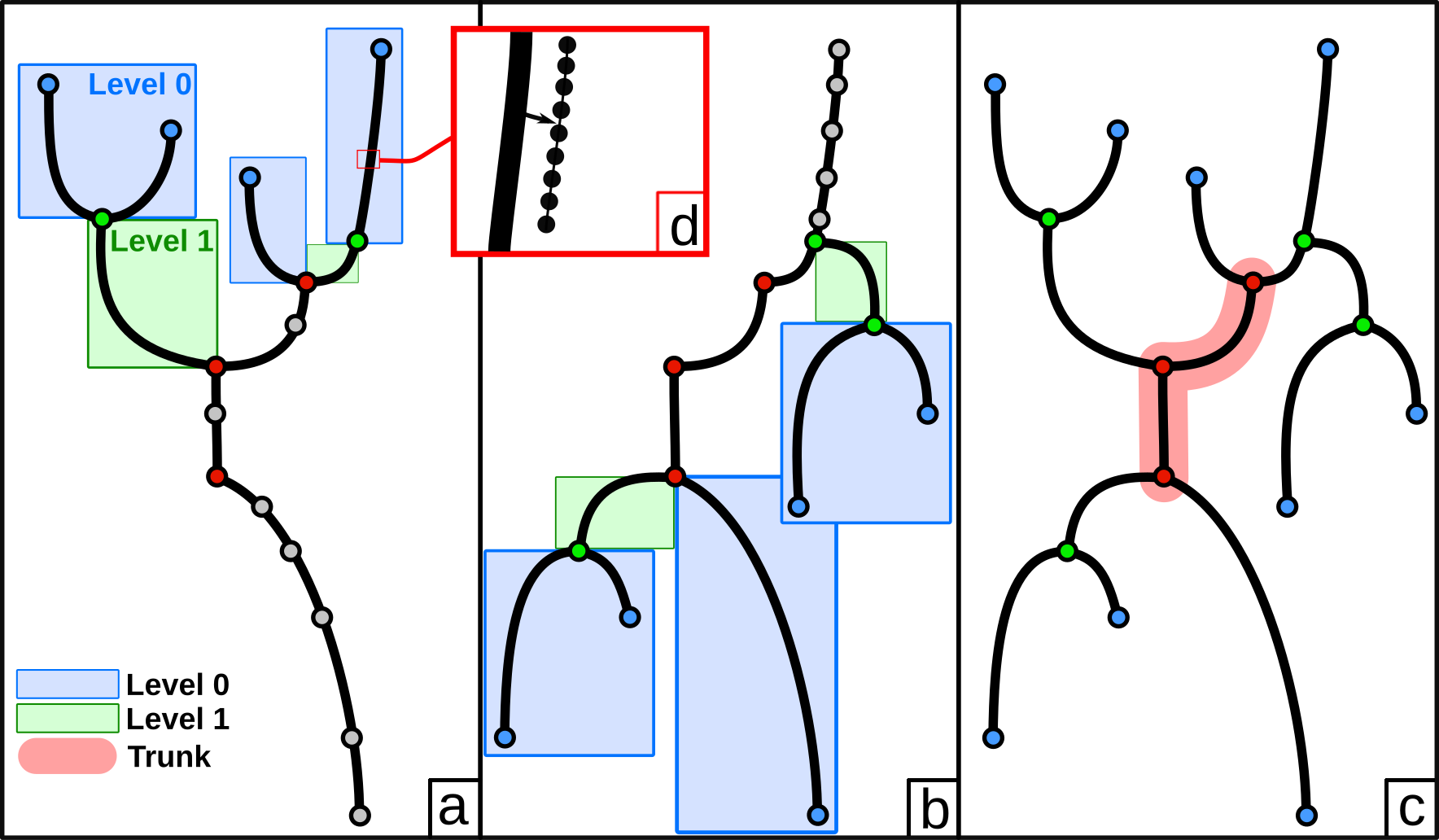}
  \vcaption{%
     A join (a) and a split (b) tree augmented with the critical nodes of the
     final tree.
     The combination of these two trees
     results in
     the final contour tree (c). The notion of \emph{level}
     (length of the shortest monotone path to the closest leaf)
     is emphasized using the blue and green boxes,
     corresponding respectively to the levels 0 and 1.
     The last monotone path can be filled using our highly parallel trunk
     procedure and is highlighted in red.
     In (d), we illustrate the list of regular vertices corresponding to
     the arc segmentation.
  }
  \label{fig:combine}
\end{figure}

For completeness we sketch here the main steps of the reference
algorithm~\cite{carr00} used to combine the join and split trees into the final
contour tree. According to this algorithm, the contour tree is created from the
two merge trees by processing their leaves one by one, adding newly created
leaves in a queue until it is empty:
\begin{enumerate}
   \item{Add leaf nodes of $\joinTree$ and $\splitTree$ to a queue
         $\combineQueue$.
      }
   \item{Pop the first node of $\combineQueue$ and add its adjacent arc in the
         final contour tree $\ct$ with its segmentation.
      }
   \item{Remove the processed node from the two trees. If this creates a new
         leaf node in the original tree, add this node into $\combineQueue$.
      }
   \item{If $\combineQueue$ is not empty, repeat from 2.}
\end{enumerate}
During phase 2, the arc and its list of regular vertices (shown in
\autoref{fig:combine}d) are processed. The list of regular vertices is visited
and all vertices not already marked are marked with the new arc identifier in
the final tree. As a vertex is both in the join and split trees, each vertex
will be visited twice. In phase 3, when a node is deleted from a merge tree,
three situations may occur. First, if the node has one adjacent arc: remove the
node along with this adjacent arc. Second, if the node has one arc up and one
down: remove the node to create a new arc which is the concatenation of the two
previous ones. Finally in all other situations, the node is not deleted yet: a
future deletion will remove it in a future iteration.

We present here a new parallel algorithm to combine the join and the split
trees, which improves the reference algorithm~\cite{carr00}. First, we define
the notion of \textit{level} of a node in a merge tree as the length of the
shortest monotone path to its closest leaf. For example, in
\autoref{fig:combine} the blue nodes are the leaves and correspond to the level
0, while the green ones at a distance of one arc correspond to the level 1.

During the combination, all the nodes and arcs at a common level can be
processed in arbitrary order. \Charles{This corresponds to the
\textbf{ArcsCombine} procedure of Alg.~\autoref{algo:ct}} We use this for
parallelism, by allowing each node (and its corresponding arc) to be processed
in parallel. Moreover, processing an arc consist of marking unvisited vertices
with an identifier. This can be done in parallel, using tasks, by processing
contiguous chunks of regular vertices. In summary, we have two nested levels of
task-parallelism available during the arc combination. First we can create
tasks to process each arc, then we can create tasks to process regular vertices
of an arc in parallel. We use this to create tasks
with a large enough computation grain size, and to avoid being constrained by
the (possible) low number of arcs to process.
In our experimental setup, we choose 10,000 vertices per task. \Charles{These
two levels of parallelism are a novelty of our approach, improving both the
load balancing and the task computation grain size tuning while also increasing
the parallelism degree.} However, we note that two synchronizations are
required. First, the procedure needs to wait for all nodes of a given level to
be processed before going to the following level. Second, data races may occur
if the node deletion is not protected in the merge trees as several nodes can
be deleted along a same arc simultaneously. A critical section is added around
the corresponding deletions. In practice, since most of the time is spent
processing arcs and their segmentations, this does not represent a performance
bottleneck.

Finally, similarly to the merge tree, there is a point where all the remaining
work is a monotone path tracing, when the contribution of the join and split
trees is reduced to one node each. We can interrupt the combination and use the
\Charles{same} trunk procedure \Charles{than} described in
\autoref{section:trunkGrowth} \Charles{for the merge tree} to process the
remaining nodes, arcs and vertices in parallel. This trunk procedure
\Charles{(corresponding to the \textbf{TrunkCombine} in Alg.~\autoref{algo:ct})}
will indeed offer a higher parallelism degree at the end of our combination
algorithm. This procedure ignores already processed vertices and project the
unvisited ones in the arcs of the remaining monotone path. Note that the size
of this trunk does not depends on the task scheduling (as it is the case for
the merge tree), but is fixed by the topology of the join and split trees.

\section{Results}
\label{section:Results}

In this section we present performance results obtained on a workstation with
two Intel Xeon E5-2630 v3 CPUs (2.4 GHz, 8 CPU cores and 16 hardware threads
each) and 64 GB of RAM.\@ By default, parallel executions will thus rely on 32
threads. These results were performed with our VTK/OpenMP based C++
implementation (provided as additional material) using \texttt{g++} version
6.4.0 and OpenMP 4.5. This implementation (called {\it Fibonacci Task-based
Contour tree}, or FTC) was built as a TTK~\cite{ttk} module. FTC uses TTK's
triangulation data structure which supports both tetrahedral meshes and regular
grids by performing an implicit triangulation with no memory overhead for the
latter. For the Fibonacci heap, we used the implementation available in Boost.

Our tests have been performed using eight data sets from various domains. The
first one, Elevation, is a synthetic data set where the scalar field
corresponds to the z coordinate, with only one connected component of level
set: the output is thus composed of only one arc. Five data sets (Ethane Diol,
Boat, Combustion, Enzo and Ftle) result from simulations and two (Foot and
Lobster) from acquisition, containing large sections of noise. For the sake of
comparison, these data sets have been re-sampled, using single floating-point
precision, on the same regular grid and have therefore the same number of
vertices.

\subsection{Merge Tree performance results}
\label{section:PerformancesMerge}


\newcommand{\jcell}[1]{#1}
\newcommand{\scell}[1]{\cellcolor[HTML]{ECF4FF}#1}


\begin{table}[t]
   \caption{%
      Running times (in seconds) of the different steps of FTC on a $512^3$
      grid for the \emph{join} and \emph{split trees} (white and gray
      backgrounds respectively). $|\mt|$ is the number of arcs in the tree.
   }
   \centering%

   \label{array:mtTime}
   \scalebox{0.765}{%
   \scriptsize%
   \begin{tabu}{%
         |l%
         {r}%
         |{r}|%
         *{5}{r}%
         |{r}|%
      }
      \toprule
                                        &                 & Sequential     & \multicolumn{5}{c|}{Parallel (32 threads on 16 cores)} & \\
                                        &                 &                &              &    Leaf      &  Arc          & Trunk         &               & \\

      Data set                                & $|\mt|$         & Overall        & Sort         & search       & growth        & growth       & Overall       & Speedup       \\
      \midrule
                                              & \jcell{1}       & \jcell{11.44}  & \jcell{0.84} & \jcell{0.14} & \jcell{0}     & \jcell{0.20} & \jcell{1.19}  & \jcell{9.57}  \\
      \multirow{-2}{*}{Elevation}             & \scell{1}       & \scell{18.71}  & \scell{0.84} & \scell{0.65} & \scell{0}     & \scell{0.20} & \scell{1.71}  & \scell{10.89} \\ \cline{1-1}
                                              & \jcell{17}      & \jcell{35.13}  & \jcell{1.31} & \jcell{0.28} & \jcell{5.16}  & \jcell{0.62} & \jcell{7.38}  & \jcell{4.75}  \\
      \multirow{-2}{*}{Ethane Diol}           & \scell{19}      & \scell{30.79}  & \scell{1.31} & \scell{0.30} & \scell{2.58}  & \scell{0.62} & \scell{4.82}  & \scell{6.38}  \\ \cline{1-1}
                                              & \jcell{5,426}   & \jcell{29.72}  & \jcell{1.24} & \jcell{0.24} & \jcell{0.07}  & \jcell{0.64} & \jcell{2.21}  & \jcell{13.41} \\
      \multirow{-2}{*}{Boat}                  & \scell{1,715}   & \scell{29.59}  & \scell{1.24} & \scell{0.40} & \scell{0.59}  & \scell{0.63} & \scell{2.88}  & \scell{10.27} \\ \cline{1-1}
                                              & \jcell{26,981}  & \jcell{37.20}  & \jcell{1.23} & \jcell{0.37} & \jcell{3.04}  & \jcell{0.61} & \jcell{5.27}  & \jcell{7.04}  \\
      \multirow{-2}{*}{Combustion}            & \scell{23,606}  & \scell{32.38}  & \scell{1.23} & \scell{0.29} & \scell{0.53}  & \scell{0.63} & \scell{2.69}  & \scell{12.03} \\ \cline{1-1}
                                              & \jcell{96,061}  & \jcell{129.62} & \jcell{1.35} & \jcell{0.36} & \jcell{12.79} & \jcell{0.69} & \jcell{15.20} & \jcell{8.52}  \\
      \multirow{-2}{*}{Enzo}                  & \scell{115,287} & \scell{43.23}  & \scell{1.35} & \scell{0.36} & \scell{4.06}  & \scell{0.77} & \scell{6.55}  & \scell{6.59}  \\ \cline{1-1}
                                              & \jcell{147,748} & \jcell{31.21}  & \jcell{1.28} & \jcell{0.37} & \jcell{0.42}  & \jcell{0.70} & \jcell{2.78}  & \jcell{11.19} \\
      \multirow{-2}{*}{Ftle}                  & \scell{202,865} & \scell{35.85}  & \scell{1.28} & \scell{0.31} & \scell{0.60}  & \scell{0.70} & \scell{2.91}  & \scell{12.31} \\ \cline{1-1}
                                              & \jcell{241,841} & \jcell{25.06}  & \jcell{1.04} & \jcell{0.26} & \jcell{0.80}  & \jcell{0.55} & \jcell{2.67}  & \jcell{9.38}  \\
      \multirow{-2}{*}{Foot}                  & \scell{286,654} & \scell{48.59}  & \scell{1.06} & \scell{0.55} & \scell{7.82}  & \scell{0.53} & \scell{9.97}  & \scell{4.87}  \\ \cline{1-1}
                                              & \jcell{472,862} & \jcell{96.34}  & \jcell{1.07} & \jcell{0.30} & \jcell{3.59}  & \jcell{0.73} & \jcell{5.71}  & \jcell{16.86} \\
      \multirow{-2}{*}{Lobster}               & \scell{490,236} & \scell{36.64}  & \scell{1.05} & \scell{0.62} & \scell{5.45}  & \scell{0.78} & \scell{7.91}  & \scell{4.62}  \\
      \bottomrule
   \end{tabu}%
}
\end{table}

\begin{figure*}
  \centering
    \includegraphics[width=1\linewidth]{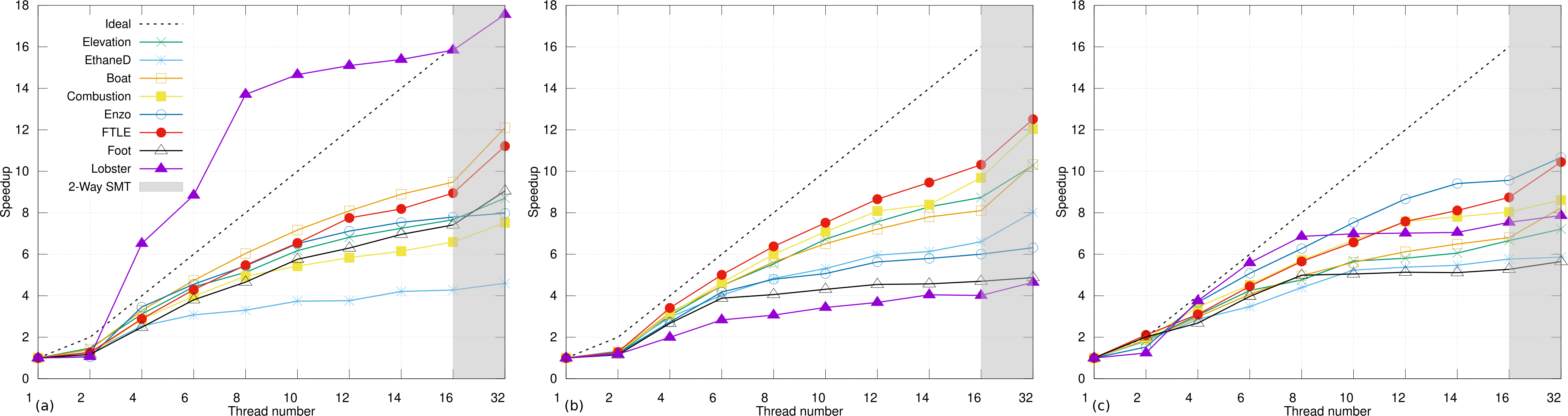}
    \vcaption{%
       FTC scalability on our $512^3$ regular grid data sets for
       (a) the join tree,
       (b) the split tree and
       (c) the contour tree
       computation.
       The gray area represents the usage of two threads
       per core with SMT (simultaneous multithreading).
    }
    \label{fig:speedup}
\end{figure*}

\autoref{array:mtTime} details the execution times and speedups of FTC for the
join and the split tree on a $512^3$ grid. One can first see that the FTC
sequential execution time varies greatly between data sets despite their equal
input size. This denotes a sensitivity on the output tree, which is common to
most merge tree algorithms. Moving to parallel executions the embarrassingly
parallel leaf search step offers very good speedups close to 14. The key step
for parallel performance is the arc growth. On most of our data sets this step
is indeed the most-time consuming in parallel, but its time varies in a large
range: this will be investigated in \autoref{section:Limitations}. The last
step is the trunk computation, which takes less than one second. Overall, with
a minimum speedup of 4.62, a maximum one of 16.86 and an average speedup of
9.29 on 16 cores, our FTC implementation achieves an average parallel
efficiency greater than 58\%. These speedups are detailed on the scaling curves
of the join and split tree computation in \autoref{fig:speedup}a and
\autoref{fig:speedup}b. The first thing one can notice is the monotonous growth
of all curves. This means that more threads always implies faster computations,
which enables us to focus on the 32-thread executions. Another interesting
point is the Lobster data set presenting speedups greater than the ideal one
when using 4 threads and more. This unexpected but welcome supra-linearity is
due to the trunk processing of our algorithm.

\begin{table}[t]
  \caption{%
     Process speed in vert/sec for the arc growth and trunk procedure in
     sequential and in parallel (join tree, grid: $512^3$).
  }%


  \centering%
\label{array:jtSpeed}
    \scalebox{1.055}{%
        \scriptsize%
        \begin{tabu}{%
                |l|%
                {r}{r}|%
                {r}{r}|%
            }
            \toprule
                & \multicolumn{2}{c|}{Sequential}  & \multicolumn{2}{c|}{Parallel}   \\
            Data set    & Arc growth  & Trunk       & Arc growth  & Trunk       \\
            \midrule
            Elevation   & \Disable{0} & 113,217,189 & \Disable{0} & 468,537,720  \\
            Ethane Diol & 472,861     & 13,862,083  & 1,003,125   & 202,175,593  \\
            Boat        & 446,981     & 13,941,128  & 933,281     & 193,274,082  \\
            Combustion  & 453,784     & 14,104,274  & 1,416,082   & 196,810,503  \\
            Enzo        & 344,129     & 11,170,128  & 2,514,479   & 138,666,543  \\
            Ftle        & 594,694     & 14,007,046  & 3,198,233   & 154,453,693  \\
            Foot        & 447,270     & 27,073,541  & 2,257,674   & 182,413,262  \\
            Lobster     & 734,705     & 19,884,438  & 2,534,264   & 135,125,845  \\
            \bottomrule
        \end{tabu}%
    }
\end{table}

As highlighted in \autoref{array:jtSpeed}, in sequential mode, the trunk step
is indeed able to process vertices much faster than the arc growth step, since
no breadth-first search traversal is performed in the trunk step (see
\autoref{section:trunkGrowth}). In parallel, the performance gap is even larger
thanks to the better parallel speedups obtained in the trunk step than in the
arc growth step. The trunk processing step is 30x faster than the arc growth in
sequential execution, and 110x faster in parallel. The arc growth is indeed 3x
faster in parallel than in sequential while the trunk is 10x faster in parallel
than in sequential. This enforces the benefits from maximizing the trunk step
in our algorithm to achieve both good performances and good speedups. However,
for a given data set, the size of the trunk highly depends on the order in
which arc growths (leaves and saddles) have been processed. Since the trunk is
detected when only one growth remains active, distinct orders in leaf and
saddle processing will yield distinct trunks of different sizes, for the same
data set. Hence maximizing the size of this trunk minimizes the required amount
of computation, especially for data sets like Lobster where the trunk
encompasses a large proportion of the domain. That is why we launch the leaf
growth tasks in the order of the scalar value of their extremum
(\autoref{section:paraLeafSearch}). Note however, that the arc growth ordering
which would maximize the size of the trunk cannot be known in advance. In a
sequential execution, it is unlikely that the  runtime will schedule the tasks
on the single thread so that the last task will be the one that corresponds to
the greatest possible trunk. Instead, the runtime will likely process each
available arc one at a time, leading to a trunk detection at the vicinity of
the root. On the contrary, in parallel, it is more likely that the runtime
environment will run out of leaves sooner, hence yielding a larger trunk than
in sequential and thus leading to increased (possibly supra-linear) speedups.

\begin{table}[t]
    \caption{%
       Stability of the execution time of FTC in parallel (join tree, 10 runs,
       $512^3$ grid).
    }

\label{array:stability512}
    \scalebox{1.2}{%
        \scriptsize%
        \centering%
        \begin{tabu}{%
                |l|%
                *{5}{r}|%
            }
            \toprule
            Data set    & Min   & Max   & Range & Average & Std.\ dev\\
            \midrule
            Elevation   & 1.17  & 1.19  & 0.02 & 1.18  & 0.01 \\
            Ethane Diol & 7.37  & 8.67  & 1.29 & 8.00  & 0.42 \\
            Boat        & 2.11  & 2.21  & 0.09 & 2.14  & 0.02 \\
            Combustion  & 4.61  & 5.27  & 0.65 & 4.89  & 0.17 \\
            Enzo        & 14.44 & 15.82 & 1.38 & 15.29 & 0.53 \\
            Ftle        & 2.75  & 2.82  & 0.07 & 2.78  & 0.02 \\
            Foot        & 2.63  & 2.70  & 0.07 & 2.67  & 0.02 \\
            Lobster     & 5.36  & 5.71  & 0.34 & 5.53  & 0.13 \\
            \bottomrule
        \end{tabu}%
    }
\end{table}

As the dynamic scheduling of the tasks on the CPU cores may vary from one
parallel execution to the next, it follows that the trunk size may also vary
across different executions, hence possibly impacting noticeably runtime
performances. As shown in \autoref{array:stability512}, the range within which
the execution times vary is clearly small compared to the average time and the
standard deviation shows a very good stability of our approach in practice.

\begin{table}[t]
   \caption{%
      \emph{Sequential} join tree computation times (in seconds) and ratios
      between libtourtre (LT), Contour Forests (CF), our preliminary
      Fibonacci Task-based Merge tree (FTM)~\cite{gueunet2017ftm} and our
      extended Fibonacci Task-based Contour tree (FTC), on a $256^3$ grid.
   }%


\label{array:jtCompareSeq}
    \scriptsize%
    \scalebox{0.86}{%
        \centering%
        \begin{tabu}{%
                |l|%
                *{1}{r}|%
                *{1}{r}|%
                *{1}{r}|%
                *{1}{r}|%
                *{3}{r}|%
            }
            \toprule
            Data set    & LT    & CF    & FTM   & FTC   & LT / FTC & CF / FTC & FTM / FTC \\
            \midrule
            Elevation   & 5.81  & 7.70  & 3.41  & 1.44  & 4.01     & 5.31     & 2.35 \\
            Ethane Diol & 11.59 & 17.75 & 7.21  & 3.61  & 3.20     & 4.91     & 1.99 \\
            Boat        & 11.84 & 17.11 & 7.81  & 3.06  & 3.86     & 5.57     & 2.54 \\
            Combustion  & 11.65 & 16.87 & 7.96  & 4.05  & 2.87     & 4.15     & 1.96 \\
            Enzo        & 14.33 & 17.99 & 18.00 & 13.62 & 1.05     & 1.32     & 1.32 \\
            Ftle        & 11.32 & 15.62 & 7.24  & 3.55  & 3.18     & 4.39     & 2.04 \\
            Foot        & 9.45  & 12.72 & 5.94  & 3.20  & 2.95     & 3.97     & 1.85 \\
            Lobster     & 11.65 & 14.80 & 14.20 & 10.05 & 1.15     & 1.47     & 1.41 \\
            \bottomrule
        \end{tabu}%
    }
\end{table}

\begin{table}[t]
   \caption{%
      \emph{Parallel} join tree computation times (in seconds) and ratios
      between libtourtre (LT~\cite{libtourte}), Contour Forests
      (CF~\cite{gueunet2016contour}), our preliminary Fibonacci
      Task-based Merge tree (FTM~\cite{gueunet2017ftm}) and our extended
      Fibonacci Task-based Contour tree (FTC) ($256^3$ grid).
   }


\label{array:jtComparePara}
    \scriptsize%
    \scalebox{0.88}{%
        \centering%
        \begin{tabu}{%
                |l|%
                *{1}{r}|%
                *{1}{r}|%
                *{1}{r}|%
                *{1}{r}|%
                *{3}{r}|%
            }
            \toprule
            Data set    & LT    & CF   & FTM  & FTC  & LT / FTC & CF / FTC & FTM / FTC \\
            \midrule
            Elevation   & 5.00  & 2.33 & 0.35 & 0.18 & 27.19    & 12.67    & 1.95 \\
            Ethane Diol & 8.95  & 4.54 & 1.24 & 0.85 & 10.52    & 5.33     & 1.46 \\
            Boat        & 8.24  & 4.40 & 0.61 & 0.29 & 28.02    & 14.96    & 2.09 \\
            Combustion  & 7.96  & 5.82 & 0.86 & 0.54 & 14.62    & 10.69    & 1.59 \\
            Enzo        & 12.18 & 8.92 & 1.91 & 1.60 & 7.60     & 5.56     & 1.19 \\
            Ftle        & 8.19  & 4.98 & 0.97 & 0.54 & 15.12    & 9.19     & 1.80 \\
            Foot        & 7.60  & 6.94 & 1.12 & 0.86 & 8.78     & 8.02     & 1.30 \\
            Lobster     & 8.40  & 9.02 & 1.69 & 0.92 & 9.03     & 9.70     & 1.82 \\
            \bottomrule
        \end{tabu}%
    }
\end{table}

Finally, in order to better evaluate the FTC performance, we
compare our approach to three reference implementations, which are, to
our knowledge, the only three public implementations supporting
augmented trees:
\begin{itemize}
   \itemsep0em
   \item{\emph{libtourtre} (LT)~\cite{libtourte}, an open source sequential
         reference implementation of the traditional algorithm~\cite{carr00};
      }
   \item{the open source implementation~\cite{ttk} of the parallel Contour
         Forests (CF) algorithm~\cite{gueunet2016contour};
      }
   \item{the preliminary version of our algorithm~\cite{gueunet2017ftm}: the
         Fibonacci Task-based Merge tree (FTM) algorithm.
      }
   \raggedbottom
\end{itemize}
In each implementation, TTK's triangulation data structure~\cite{ttk} is used
for mesh traversal. Due to its important memory consumption, we were unable to
run CF on the $512^3$ data sets on our workstation. Thus, we have created a
smaller grid ($256^3$ vertices) with down-sampled versions of the scalar fields
used previously. For the first step of this comparison we are interested in the
sequential execution. The corresponding results are reported in
\autoref{array:jtCompareSeq}. We note that in sequential, Contour Forests and
libtourtre implements the same algorithm. Our sequential implementation is
about 3.90x faster than Contour Forests and more than 2.70x faster than
libtourtre for most data sets. This is due to the faster processing speed of
our trunk step. Thanks to our fine grain optimizations, computing a merge tree
using FTC is faster than with FTM, by a factor of 1.93x on average in our test
cases.
\Charles{As far as this improvement is concerned}, up to 40\% is due to the
early trunk detection for data sets with large arcs (cf.
\autoref{sec:saddleGrowthSec}), 40\% to the lazy valence computation (cf.
\autoref{section:paraSaddleStop}) and 20\% is due to the use of an array of
structure (cf. \autoref{section:Parallel}). The parallel results for the merge
tree implementation are presented in \autoref{array:jtComparePara}. The
sequential libtourtre implementation starts by sorting all the vertices, then
computes the tree. Using a parallel sort instead of the serial one is
straightforward. Thus, we used this naive parallelization of LT in the results
reported in \autoref{array:jtComparePara} with 32 threads. As for Contour
Forests we report the best time obtained on the workstation, which is not
necessarily with 32 threads. Indeed, as detailed in~\cite{gueunet2016contour}
increasing the number of threads in CF can result in extra work due to
additional redundant computations. This can lead to greater computation times,
especially on noisy data sets. The optimal number of threads for CF has thus to
be chosen carefully. On the contrary, FTM and FTC always benefit from the
maximum number of hardware threads. In the end, FTC largely outperforms the
other implementations for all data sets: libtourtre by a factor 15.11x (in
average), Contour Forests by a factor 9.51x (in average) and our preliminary
approach \cite{gueunet2017ftm} by a factor of 1.63x. We emphasize that the two
main performance bottlenecks of CF in parallel, namely extra work and load
imbalance, do not apply to FTC thanks to the arc growth algorithm and to the
dynamic task scheduling.

\subsection{Contour Tree performance results}
\label{section:PerformancesContour}

\begin{table}[t]
  \caption{%
      Contour tree computation times (in seconds) with FTC on the $512^3$ grid.
      Extremum detection is reported under the Leaf Search column. The
      concurrent computation of the two merge trees is reported under the MT
      column. The parallel combination of these trees is in the Combine column.
    }


\label{array:ctTime}
    \scalebox{0.735}{%
        \scriptsize%
        \centering%
        \begin{tabu}{%
                |l%
                {r}%
                |{r}|%
                *{5}{r}%
                |{r}|%
            }
            \toprule
                        &         & Sequential& \multicolumn{5}{c|}{Parallel (32 threads on 16 cores)}         &             \\
                        &         &           & \multicolumn{5}{c|}{}                                          &             \\
            Data set    & $|\mt|$ & Overall & Sort & Leaf search & MT    & Combine & Overall & Speedup \\
            \midrule
            Elevation   & 1       & 20.92   & 1.07 & 0.61        & 1.08  & 0       & 2.77    & 7.54    \\
            Ethane Diol & 35      & 70.63   & 1.48 & 0.44        & 9.29  & 0.61    & 11.84   & 5.96    \\
            Boat        & 7,140   & 59.33   & 1.39 & 0.48        & 2.55  & 2.78    & 7.21    & 8.22    \\
            Combustion  & 50,586  & 76.00   & 1.37 & 0.49        & 5.22  & 1.57    & 8.66    & 8.76    \\
            Enzo        & 211,346 & 215.08  & 1.47 & 0.58        & 15.63 & 1.99    & 19.68   & 10.92   \\
            Ftle        & 350,602 & 73.42   & 1.46 & 0.56        & 3.32  & 1.73    & 7.08    & 10.36   \\
            Foot        & 528,494 & 83.44   & 1.15 & 0.77        & 10.06 & 3.01    & 14.99   & 5.56    \\
            Lobster     & 963,068 & 143.15  & 1.21 & 0.89        & 9.80  & 6.77    & 18.68   & 7.66    \\
            \bottomrule
        \end{tabu}%
    }
\end{table}

\begin{figure}
    \centering
    \includegraphics[width=1\linewidth]{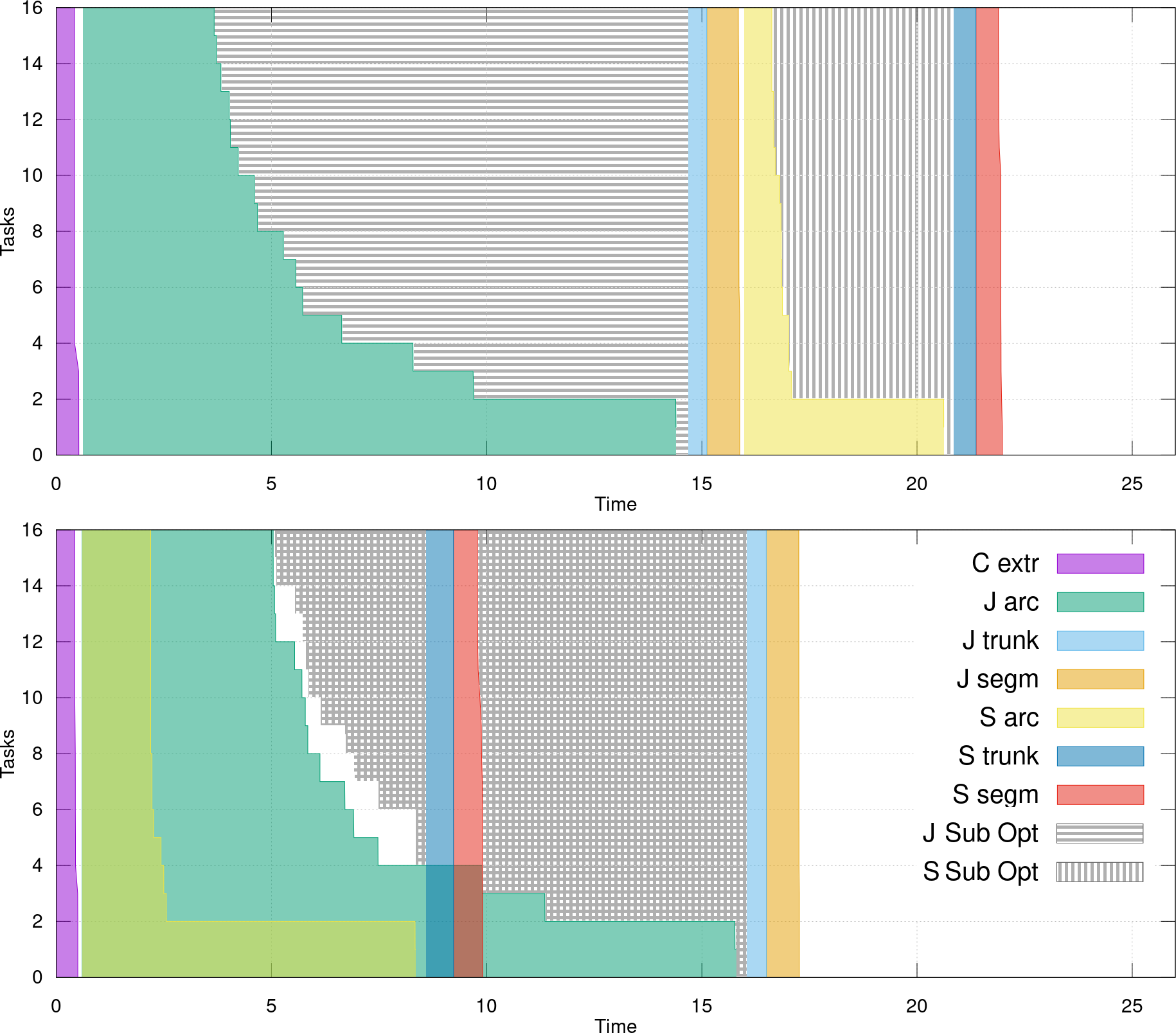}
    \vcaption{%
       Number of remaining tasks over time for a parallel execution on the Enzo
       data set. Each step of the algorithm is shown with a distinct color. The
       suboptimal sections are shown with areas stripped in gray. At the top,
       the join and split trees are computed separately (join tree first). At
       the bottom, they are computed concurrently (hence, at a given time, the
       number of remaining tasks is the sum of the overlapping curves).
    }
    \label{fig:tasksCTboth}
\end{figure}

\autoref{array:ctTime} details execution times for our contour tree
computation. As for the merge tree, the sequential times vary across data sets
due to the output sensitivity of the algorithm. A single leaf search is
performed for both merge trees (corresponding to a 25\% performance improvement
for this step, both in sequential and in parallel).
\Charles{Focusing on} parallel executions, most of the time is spent computing
the join and the split trees as reported under the MT column. We further
investigate this step later with \autoref{array:ctSchedule} and
\autoref{fig:tasksCTboth}. As for the combination, it takes longer to compute
for larger trees, with the exception of the Boat data set having a particularly
small trunk. This illustrates the output sensitivity of our combination
algorithm, as detailed in \autoref{array:ctCombine}. Our contour tree
computation algorithm results in speedups varying between 5.56 and 10.92 in our
test cases, with an average of 8.12 corresponding to an average parallel
efficiency of 50.75\%.

The evolution of these speedups as a function of the number of threads is shown
in \autoref{fig:speedup}c. These speedups are consistent with those of the
merge tree (\autoref{fig:speedup}a and \autoref{fig:speedup}b). Our algorithm
benefits from the dynamic task scheduling and its workload does not increase
with the number of threads. This also applies to the our  combination
algorithm. Therefore in theory, the more threads are available, the faster FTC
should compute the contour tree. In practice, this translates into
monotonically growing curves as shown in \autoref{fig:speedup}. For the contour
tree computation, curves shown \autoref{fig:speedup}c have lower slopes than
those of the merge trees (\autoref{fig:speedup}a and \autoref{fig:speedup}b).
This is mainly due to the combination procedure which has a smaller speedup
than our merge tree procedure as detailed below in \autoref{array:ctCombine}.

\begin{table}[t]
    \centering%
    \caption{%
       Merge tree processing time during the parallel contour tree computation
       ($512^3$ grid). \emph{JT then ST} reports results obtained by separately
       computing first the join tree then the split tree, leading to the
       successive execution of two distinct suboptimal sections. In \emph{Task
       overlapping}, the two trees are concurrently computed and overlap occurs
       in their task scheduling.
    }
    \scalebox{1.09}{%
        \scriptsize%
        \label{array:ctSchedule}
        \begin{tabu}{%
                |l|%
                {r}|%
                {r}|%
                {r}|%
                {r}{r}|%
            }
            \toprule
            Data set    & JT then ST & Task overlapping & Overlap speedups \\
            \midrule
            Elevation   & 2.25       & 1.73             & 1.30 \\
            Ethane Diol & 12.80      & 10.14            & 1.26 \\
            Boat        & 3.90       & 3.11             & 1.25 \\
            Combustion  & 6.49       & 5.55             & 1.17 \\
            Enzo        & 21.34      & 17.69            & 1.21 \\
            Ftle        & 4.74       & 3.86             & 1.23 \\
            Foot        & 12.14      & 10.48            & 1.16 \\
            Lobster     & 14.45      & 10.81            & 1.34 \\
            \bottomrule
        \end{tabu}%
    }
\end{table}

\noindent \textbf{Task overlapping.} \label{par:TasksScheduling}
\autoref{array:ctSchedule} presents speedups obtained by computing both trees
concurrently, allowing tasks to overlap in their scheduling during the merge
tree parallel computation, thanks to the complete taskification of our
implementation. This overlap reduces the size of the suboptimal section, as
shown in \autoref{fig:tasksCTboth}. This strategy results in speedups up to
1.34x (1.24x in average) compared to a successive computation of the two trees.

Indeed, as mentioned in \autoref{section:treeMix}, during the arc growth
computation, the number of remaining tasks becomes smaller than the number of
threads. As illustrated \autoref{fig:tasksCTboth} this leads to a suboptimal
section, where  some available threads are left idle. On this chart, the
suboptimal section is shown using the stripped gray area. If the join and split
trees are computed one after the other, (\autoref{fig:tasksCTboth}, top chart)
we observe two distinct suboptimal sections: one for the join tree and one for
the split tree. In contrast, when the join and split trees are computed
simultaneously (\autoref{fig:tasksCTboth}, bottom chart) the OpenMP runtime can
pick tasks among either trees, hence reducing the area of the stripped section.
Moreover, at the bottom chart of \autoref{fig:tasksCTboth}, when the arc growth
procedure of the split tree finishes, that of the join tree is still
processing. The remaining steps of the split tree computation (trunk processing
and regular vertex segmentation) continues in the meantime, which contributes
to reducing the suboptimal section (blue and red columns in
\autoref{fig:tasksCTboth}). At the end, this task overlapping strategy results
in a smaller stripped area and so in an improved parallel efficiency. In the
same manner, the total time of the leaf search plus merge tree computation
reaches 21.34 seconds when merge trees are computed one after the other and
17.69 seconds in an overlapped merge trees execution
(cf.~\autoref{array:ctSchedule}).

\begin{table}[t]
    \centering%
    \caption{%
       Combination procedure times for sequential and parallel executions with
       and without the trunk processing, compared to the sequential combination
       procedure of our preliminary version (FTM~\cite{gueunet2017ftm}, $512^3$
       grid). \Charles{The 0 values for the Elevation data-set are due to the
       filiform nature of its merge trees (which implies instantaneous
combinations).}
    }%
    \scalebox{0.915}{%
        \scriptsize%
        \label{array:ctCombine}
        \begin{tabu}{%
                |l|%
                {r}|%
                {r}{r}|%
                {r}{r}|%
                {r}|%
            }
            \toprule
                                &                               & \multicolumn{2}{c|}{Sequential}
                                & \multicolumn{2}{c|}{Parallel} & FTM / parallel \\
            Data set            & FTM                           & no trunk    & trunk
                                & no trunk                      & trunk       & FTC + trunk \\ 
            \midrule
            \Disable{Elevation} & \Disable{0}                   & \Disable{0} & \Disable{0} & \Disable{0} & \Disable{0} & \Disable{N.A.} \\
            Ethane Diol         & 3.23                          & 3.82        & 6.40        & 2.51        & 0.54        & 5.98 \\ 
            Boat                & 3.11                          & 3.99        & 3.60        & 2.63        & 2.64        & 1.17 \\ 
            Combustion          & 3.29                          & 3.63        & 5.62        & 3.30        & 1.49        & 2.20 \\ 
            Enzo                & 4.72                          & 4.52        & 7.03        & 4.18        & 1.90        & 2.48 \\ 
            Ftle                & 4.79                          & 5.13        & 7.62        & 5.01        & 1.70        & 2.81 \\ 
            Foot                & 4.63                          & 4.46        & 5.15        & 5.04        & 3.14        & 1.47 \\ 
            Lobster             & 7.11                          & 7.22        & 7.46        & 8.33        & 6.72        & 1.05 \\ 
            \bottomrule
        \end{tabu}%
    }
\end{table}


\noindent\textbf{Parallel Combination.} \label{par:ParallelCombine} For the
combination step, we report in \autoref{array:ctCombine} comparisons between
various versions of our algorithm and the reference algorithm~\cite{carr00}
implemented in our preliminary approach~\cite{gueunet2017ftm}. Note that our
parallel algorithm executed sequentially, without triggering the fast trunk
procedure, corresponds to the reference sequential algorithm~\cite{carr00}.
According to this table, enabling the trunk on a sequential execution of our
new algorithm is slower by 33\% in average. We believe this is due to two
reasons. First, each regular vertex additionally checks if it should be added
to the current arc (\autoref{section:paraCombination}). Second, the trunk
procedure may re-visit some vertices already visited by the arc combination
procedure, which results in redundant visits
(\autoref{section:paraCombination}). In our test cases, this redundant work
affects less than 1\% of the total number of vertices. In contrast, enabling
the trunk procedure in a parallel execution is necessary to achieve significant
speedups, by an average factor of 1.98x in \autoref{array:ctCombine}, with
respect to the sequential reference algorithm implemented in FTM.\@ Indeed, in
the parallel combination algorithm the number of arcs at each level decreases,
inducing a decreasing trend in the number of vertices processed (and tasks
created) at each leave, and leading to another suboptimal section. The trunk
procedure occurs at a point where the arcs combination is likely to use a small
number of tasks and replace it by a highly parallel processing, thus improving
parallel efficiency. Finally, according to these observations, we choose to
trigger the trunk processing only for parallel executions.


\begin{table}[t]
    \caption{%
      \emph{Sequential} contour tree computation times (in seconds) and ratios
      between libtourtre (LT~\cite{libtourte}), Contour Forests
      (CF~\cite{gueunet2016contour}), our preliminary Fibonacci Task-based
      Merge tree~\cite{gueunet2017ftm} adapted for Contour trees (FTM-CT) and
      our current Fibonacci Task-based Contour tree FTC, on a $256^3$ grid.
    }%

\label{array:ctCompareSeq}
    \scriptsize%
    \scalebox{0.795}{%
        \centering%
        \begin{tabu}{%
                |l|%
                *{1}{r}|%
                *{1}{r}|%
                *{1}{r}|%
                *{1}{r}|%
                *{3}{r}|%
            }
            \toprule
            Data set    & LT    & CF    & FTM-CT & FTC   & LT / FTC & CF / FTC & FTM-CT / FTC \\
            \midrule
            Elevation   & 10.84 & 8.15  & 6.25   & 2.82  & 3.83     & 2.88     & 2.21 \\
            Ethane Diol & 21.54 & 17.73 & 12.06  & 6.61  & 3.25     & 2.67     & 1.82 \\
            Boat        & 21.10 & 16.63 & 12.55  & 5.68  & 3.71     & 2.92     & 2.20 \\
            Combustion  & 21.52 & 16.92 & 13.35  & 7.38  & 2.91     & 2.29     & 1.80 \\
            Enzo        & 27.79 & 19.71 & 26.23  & 19.33 & 1.43     & 1.01     & 1.35 \\
            Ftle        & 23.05 & 15.89 & 13.17  & 7.33  & 3.14     & 2.16     & 1.79 \\
            Foot        & 19.24 & 13.41 & 14.25  & 9.77  & 1.96     & 1.37     & 1.45 \\
            Lobster     & 23.39 & 51.32 & 22.96  & 17.04 & 1.37     & 3.01     & 1.34 \\
            \bottomrule
        \end{tabu}%
    }
\end{table}

\noindent\textbf{Comparison.} \label{section:CTComparison}
For the contour tree computation we compare our approach with the three public
reference implementations computing the augmented contour tree. Results are
shown in \autoref{array:ctCompareSeq}. Due to the important memory consumption
of Contour Forests~\cite{gueunet2016contour}, we were unable to run these tests
on our $512^3$ regular grid. Results are reported using a down-sampled $256^3$
grid. Our implementation in sequential mode outperforms the t\Charles{h}ree
others for every data set. FTC is in average 2.70x faster than libtourtre and
2.29x faster than Contour Forests. In sequential, these two implementations
correspond to the reference algorithm~\cite{carr00}. As shown with the merge
tree in \autoref{section:PerformancesMerge}, our algorithm is able in practice
to process vertices faster thanks to the trunk step, hence the observed
improvement. FTC is also 1.75x faster than FTM thanks to the fine grain
optimizations introduced in this paper.

\begin{table}[t]
    \caption{%
      \emph{Parallel} contour tree computation times (in seconds) and ratios
      between libtourtre (LT~\cite{libtourte}), Contour Forests
      (CF~\cite{gueunet2016contour}), our preliminary Fibonacci Task-based
      Merge tree~\cite{gueunet2017ftm} adapted for Contour Tree (FTM-CT) and
      our current Fibonacci Task-based Contour trees (FTC), on a $256^3$ grid.
    }%
\label{array:ctComparePara}
    \scriptsize%
    \scalebox{0.81}{%
        \centering%
        \begin{tabu}{%
                |l|%
                *{1}{r}|%
                *{1}{r}|%
                *{1}{r}|%
                *{1}{r}|%
                *{3}{r}|%
            }
            \toprule
            Data set    & LT    & CF   & FTM-CT & FTC  & LT / FTC & CF / FTC & FTM-CT / FTC \\
            \midrule
            Elevation   & 5.00  & 2.33 & 0.73   & 0.40 & 12.31    & 5.73     & 1.79      \\
            Ethane Diol & 8.95  & 4.54 & 2.13   & 1.23 & 7.24     & 3.67     & 1.72      \\
            Boat        & 8.24  & 4.40 & 1.39   & 0.92 & 8.93     & 4.77     & 1.50      \\
            Combustion  & 7.96  & 5.82 & 1.73   & 1.15 & 6.86     & 5.01     & 1.49      \\
            Enzo        & 12.18 & 8.92 & 3.90   & 2.87 & 4.23     & 3.09     & 1.35      \\
            Ftle        & 8.19  & 4.98 & 2.55   & 1.35 & 6.03     & 3.66     & 1.87      \\
            Foot        & 7.60  & 6.94 & 4.38   & 3.10 & 2.44     & 2.23     & 1.40      \\
            Lobster     & 8.40  & 9.02 & 6.86   & 4.66 & 1.80     & 1.93     & 1.47      \\
            \bottomrule
        \end{tabu}%
    }
\end{table}

For the comparison in parallel, results are presented in
\autoref{array:ctComparePara}. For libtourtre, a naive parallelization is
achieved by using the GNU parallel sort and by computing the two merge trees
concurrently. For contour forests, we present the best time using the optimal
number of threads (not necessarily 32). Again, FTC is the fastest for all our
test cases. It outperforms libtourtre by an average factor of 6.23x (up to
8.93x for real-life data sets), our naive parallelization of libtourtre having
a maximum speedup of 2.81x on 16 cores. FTC is also faster than Contour Forests
by a factor 3.76x, taking benefits from the dynamic task scheduling and from
the absence of additional work in parallel. Finally, FTC outperforms FTM by a
factor 1.58x in average on our data sets thanks to our fine grain
optimizations, to our task overlapping strategy for the merge trees, and to the
parallel combination, all introduced on this paper.

\subsection{Limitations}
\label{section:Limitations}

The main limitation of the preliminary version of our
work~\cite{gueunet2017ftm} is the presence of so-called suboptimal sections.
Two contributions presented in this paper aim at reducing their performance
impact. First, the early trunk detection stops the arc growth processing
sooner, increasing the trunk size, as explained in
\autoref{section:trunkGrowth}. Second, for the contour tree computation, tasks
of both merge trees are created concurrently (allowing them to overlap), thus
reducing the suboptimal section (cf. \autoref{fig:tasksCTboth}).

We have considered using task priorities to maximize the task overlapping, or
to minimize the suboptimal sections. We have first studied simple heuristics
(based e.g.\ on the higher number of extrema) to choose which tree will be
computed with the high task priority (\autoref{section:treeMix}). However no
simple heuristic led to the best choice for all our data sets. We thus
arbitrarily assign the high priority to the split tree tasks. Second, we have
also considered using task priorities to maximize the number of active tasks at
the end of the arc growth step. However this would likely reduce the trunk
size, which would lead to lower overall performance results since the trunk
processing is two orders of magnitude faster than the arc growth one
(\autoref{section:PerformancesMerge}). Finally, we have also tried using
distinct task priorities for the successive steps of our algorithm (and still
for the two merge trees), but to no avail.

\begin{figure}
    \centering
    \includegraphics[width=1\linewidth]{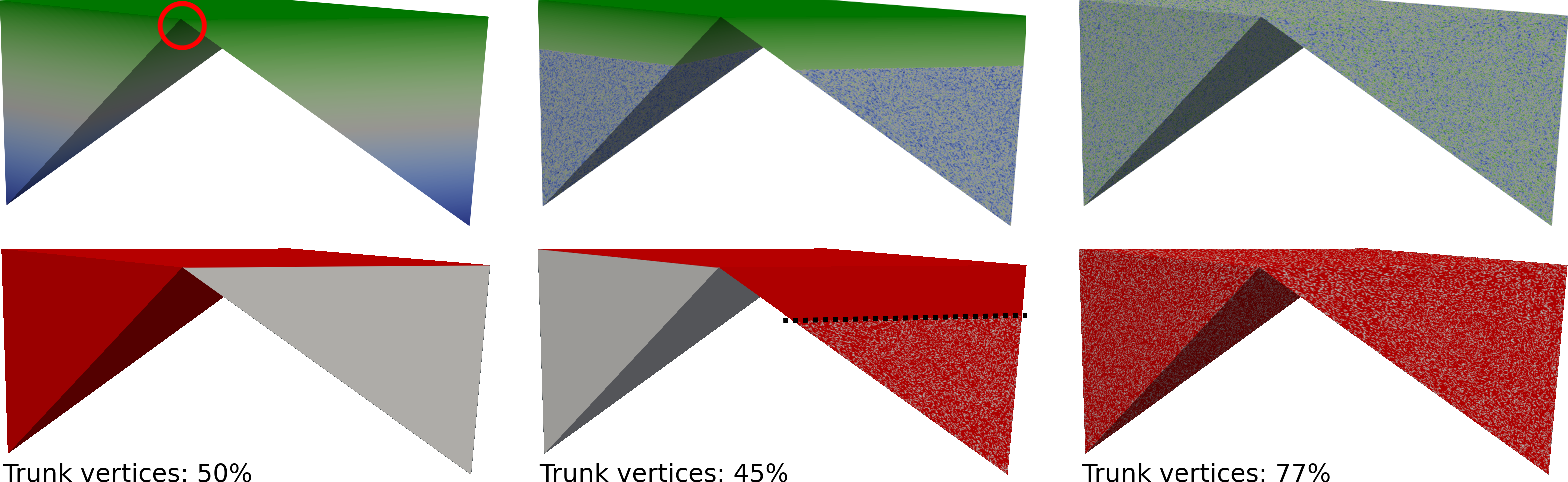}
    \vcaption{%
       Worst case data set with the initial scalar field (top left, blue to
       green), with 50\% (top middle), and with 100\% of randomness (top
       right). The red circle indicates a saddle point induced by the Elevation
       scalar field, called hereafter ``natural saddle''. Vertices processed by
       the trunk procedure are shown in red (bottom).
    }
\label{fig:worstFig}
\end{figure}

\begin{figure}
    \centering
    \includegraphics[width=1\linewidth]{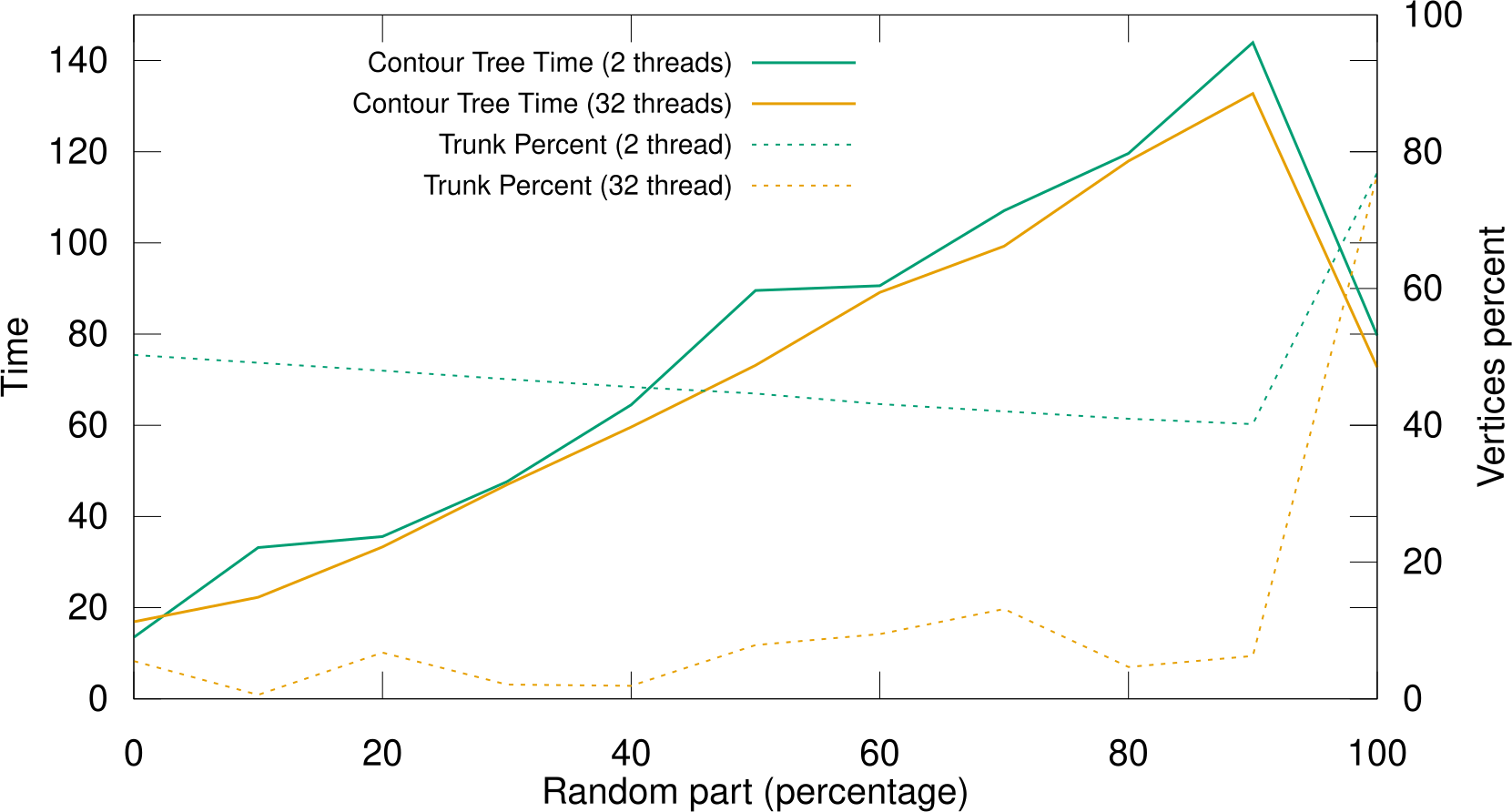}
    \vcaption{%
       FTC contour tree computation time for 2 and 32 threads on our worst case
       data set as the random part progresses form 0 to 100\% (plain lines,
       left axis) and percentage of vertices processed by the trunk procedure
       (dashed lines, right axis).
    }
\label{fig:worstTime}
\end{figure}

In order to   illustrate the performance impact of these suboptimal sections,
we have created a worst case data set composed of only two large arcs as
illustrated on the left of \autoref{fig:worstFig}. As expected, the speedup of
the join tree arc growth step on this data set does not exceed 2, even when
using 32 threads (results not shown). Then we randomize this worst case data
set gradually, starting by the leaf side as illustrated in
\autoref{fig:worstFig} and report the corresponding contour tree computation
times with 2 and 32 threads in \autoref{fig:worstTime}. As the random part
progress (from 0 to 90\%) the execution time increases. This is due to the
output sensitive nature of contour tree algorithms, but also to the smaller
trunk size when the percentage of random vertices increase.
\autoref{fig:worstFig} shows the vertices processed by the trunk procedure (in
red, bottom) for different percentages of randomness. Increases in the level of
randomness (from left to right) decrease the number of vertices processed by
the efficient trunk procedure. When the level of randomness goes beyond the
\emph{natural saddle} of the data set (red circle, \autoref{fig:worstFig}), the
specifically designed 2-arc worst-case structure disappears and the data set
becomes similar to a fully random data set. Interestingly, such a random data
set is no longer the worst case for our algorithm (see the execution time drop
at 100\%, \autoref{fig:worstTime}), as the set of vertices processed by the
efficient trunk procedure remains sufficiently large (\autoref{fig:worstFig},
right).

\section{Application}
\label{section:Application}

\begin{figure}
   \hfill
   \centering
   \subfloat[]{%
      \includegraphics[width=0.31\linewidth]{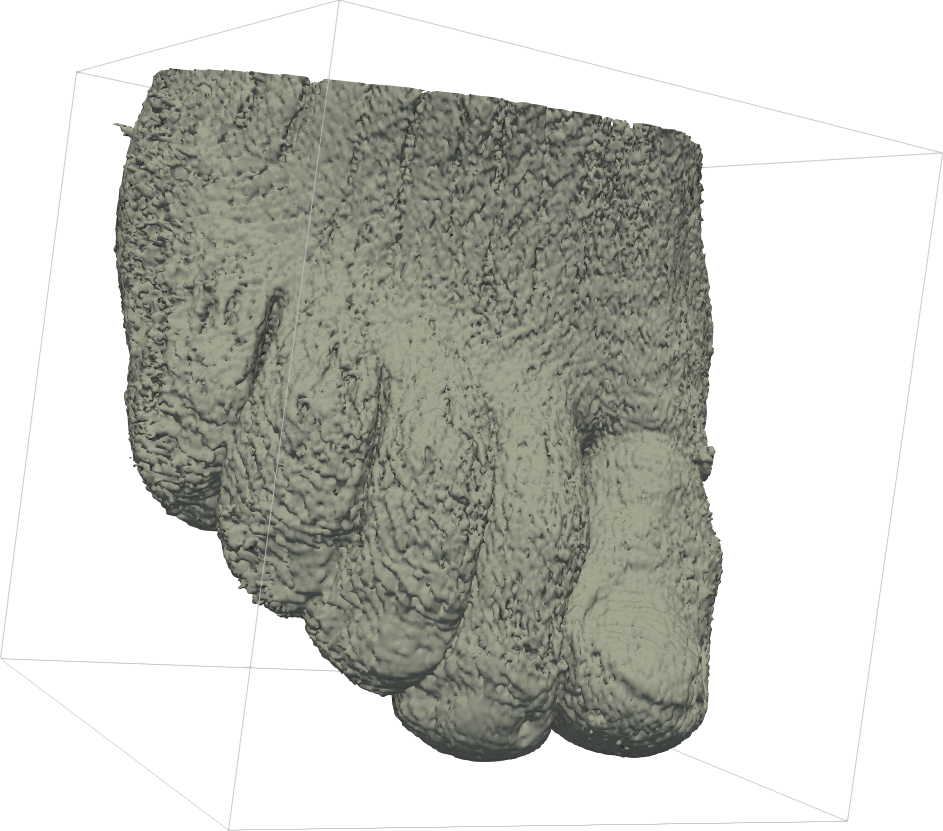}
      \label{fig:appliFoot}
   }
   \hfill
   \centering
   \subfloat[]{%
      \includegraphics[width=0.31\linewidth]{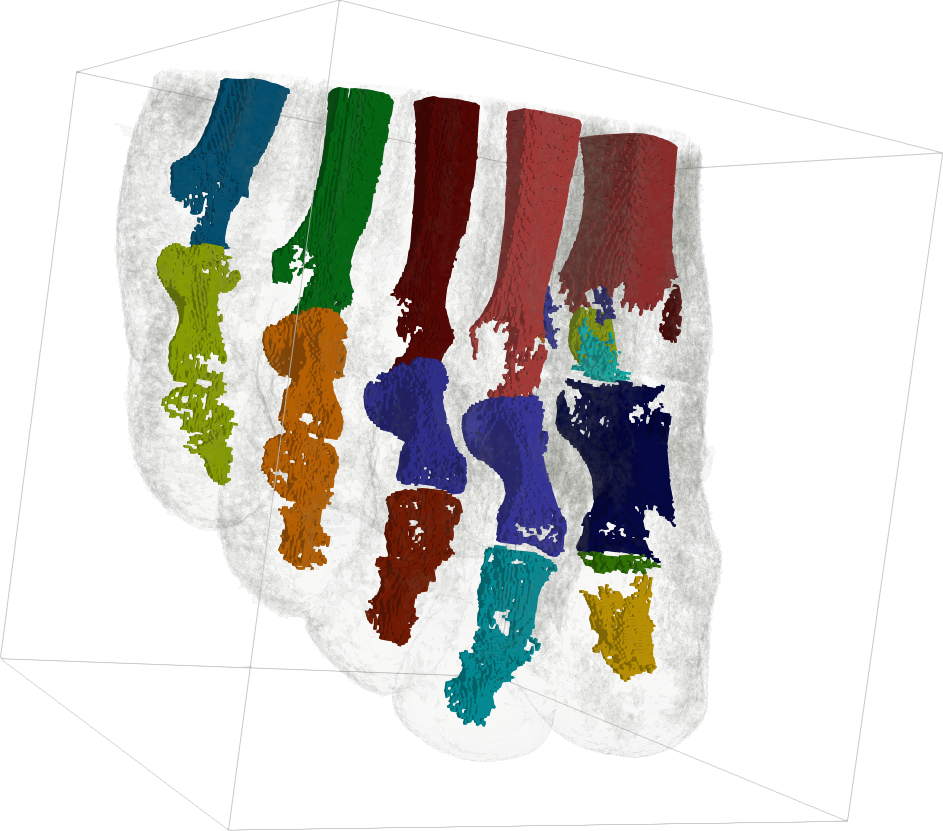}
      \label{fig:appliPhalange}
   }
   \centering
   \subfloat[]{%
      \includegraphics[width=0.31\linewidth]{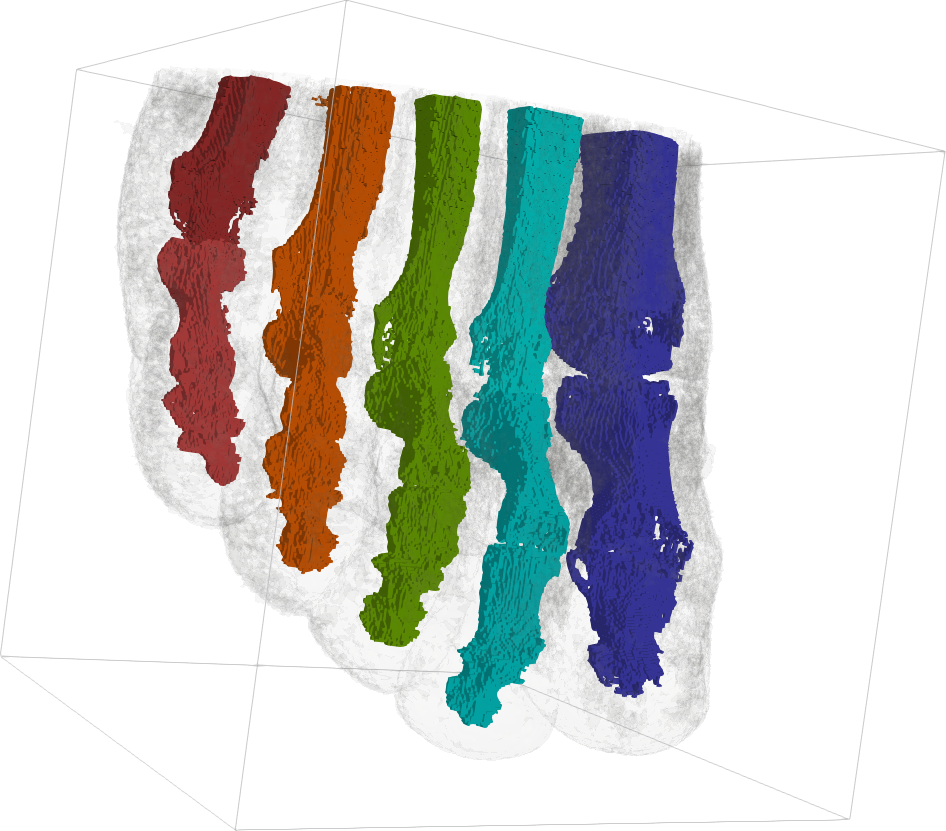}
      \label{fig:appliFinger}
   }
   \vcaption{%
      The Foot data set is a 3D scan of a human foot on which the scalar
      field is the density. We use the split tree segmentation
      to extract bones.
      (a) One contour corresponding to the skin of the foot.
      (b) The different bones highlighted using the segmentation of the deepest
      arcs of the tree.
      (c) Using topological simplification enables us to identify bones
      belonging to a same toe.
   }
   \label{fig:appli}
\end{figure}

The merge tree is a well known tool for data segmentation used in various
applications. It is especially used in the medical domain~\cite{carr04} as
illustrated by \autoref{fig:appli} which shows a 3D scan of a human foot. The
scalar field is the matter density, different densities corresponding to
different tissues. The skeleton is easy to detect as it corresponds to the
highest density. We can extract the corresponding regions using the
segmentation of the deepest arcs of the split tree (the arcs adjacent to the
leaves) as shown in \autoref{fig:appli}b. By using topological simplification
we can merge regions of interest to identify bones belonging to the same toe as
illustrated by \autoref{fig:appli}c. Thanks to our approach this processing can
now be done in a handful of seconds, even for $512^3$ grids.
\Charles{In particular, the 10x speedups obtained by our approach over a
sequential execution (\autoref{array:mtTime})
have been shown to be highly relevant for such interactive data exploration
tasks~\cite{Niels}.}

\section{Conclusion}

In this paper, we have presented a new algorithm to compute both the augmented
merge and contour trees on shared-memory multi-core architectures. This new
approach makes use of the Fibonacci heaps to completely revisit the traditional
algorithm and compute the contour tree using independent local growths which
can be expressed using tasks. This work is well suited for both regular grids
and unstructured meshes. We also provided a lightweight generic VTK-based C++
reference implementation of our approach, based on the OpenMP task runtime.
This implementation is the fastest to our knowledge to compute these
topological data structures in augmented mode, both sequentially and in
parallel. Moreover, we presented a task overlapping strategy obtained thanks to
the complete taskification of our implementation, as well as fine grain
optimizations and a novel parallel algorithm for the combination of the join
and split trees into the output contour tree. This makes our overall approach
clearly outperform previous work in all our test cases.

As future work, we plan to extend our approach in two different ways.
While our efforts focused so far on time efficiency, we would like to further
improve the memory footprint of our implementation, to be able to address
significantly larger data sets.
We also believe that our task-based approach may be especially relevant for
\textit{in-situ} visualization, where the analysis code is executed in parallel
and in synergy with the simulation code generating the data.



\vspace{-1ex}
\ifCLASSOPTIONcompsoc%
  \section*{Acknowledgments}
\else
  \section*{Acknowledgment}
\fi

{\vspace{-1ex}
\footnotesize
   This work is partially supported by the BPI grant ``AVIDO'' (Programme
   d'Investissements d'Avenir, reference P112017-2661376/DOS0021427) and by
   the ANRT, in the
   framework of the CIFRE partnership between Sorbonne Universit\'e and Kitware
   SAS CIFRE (reference \#2015/1039).
   \Charles{The authors would like to thank the anonymous reviewers for their
   thoughtful remarks and suggestions.}
}
\vspace{-1ex}

\ifCLASSOPTIONcaptionsoff%
  \newpage
\fi



\bibliographystyle{IEEEtran}
\bibliography{paper}





\end{document}